\documentclass[aps,prl,reprint, amsmath, amssymb,superscriptaddress]{revtex4-1}
 
\usepackage{bm}
\usepackage[retainorgcmds]{IEEEtrantools}
\usepackage{graphicx}
\usepackage{mathrsfs}
\usepackage{amsmath}
\usepackage{amsfonts}
\usepackage{amssymb}
\usepackage{color}
\usepackage{times,txfonts}
\usepackage{nicefrac}
\usepackage{ragged2e}
\usepackage{tikz}
\usepackage{bbm}
\usepackage{mathtools}

\usepackage[colorlinks=true,linkcolor=blue,urlcolor=blue,citecolor=blue,pdfusetitle]{hyperref}

\usepackage{blindtext}

\newcommand{\tr}{\mathrm{tr}}

\makeatletter
\newsavebox{\@brx}
\newcommand{\llangle}[1][]{\savebox{\@brx}{\(\m@th{#1\langle}\)}%
  \mathopen{\copy\@brx\kern-0.5\wd\@brx\usebox{\@brx}}}
\newcommand{\rrangle}[1][]{\savebox{\@brx}{\(\m@th{#1\rangle}\)}%
  \mathclose{\copy\@brx\kern-0.5\wd\@brx\usebox{\@brx}}}
\makeatother

\newcommand{\id}{\mathbbm{1}}
\newcommand{\rhossV}{|\rho_{\rm ss}\rrangle}
\newcommand{\idV}{\llangle \id |}
\newcommand{\rhoss}{\rho_{\rm ss}}

\newcommand{\size}{r}

\begin{document}

\title{Patterns in the jump-channel statistics of open quantum systems}
\date{\today}
\author{Gabriel T. Landi}
\email{glandi@ur.rochester.edu}
\affiliation{Department of Physics and Astronomy, University of Rochester, Rochester, New York 14627, USA}

\begin{abstract}

A continuously measured quantum system with multiple jump channels gives rise to a stochastic process described by random jump times and random emitted symbols, representing each jump channel. 
While much is known about the waiting time distributions, very little is known about the statistics of the emitted symbols. 
In this letter we fill in this gap. 
First, we provide a full characterization of the resulting stochastic process, including efficient ways of simulating it, as well as determining the underlying memory structure. 
Second, we show how to unveil patterns in the stochastic evolution: 
Some systems support closed patterns, wherein the evolution runs over a finite set of states, or at least recurring states.
But even if neither is possible, we show that one may still cluster the states approximately, based on their ability to predict future outcomes. 
We illustrate these ideas by studying transport through a boundary-driven one-dimensional XY spin chain.


\end{abstract}

\maketitle{}

\begin{figure}
    \centering
    \includegraphics[width=0.45\textwidth]{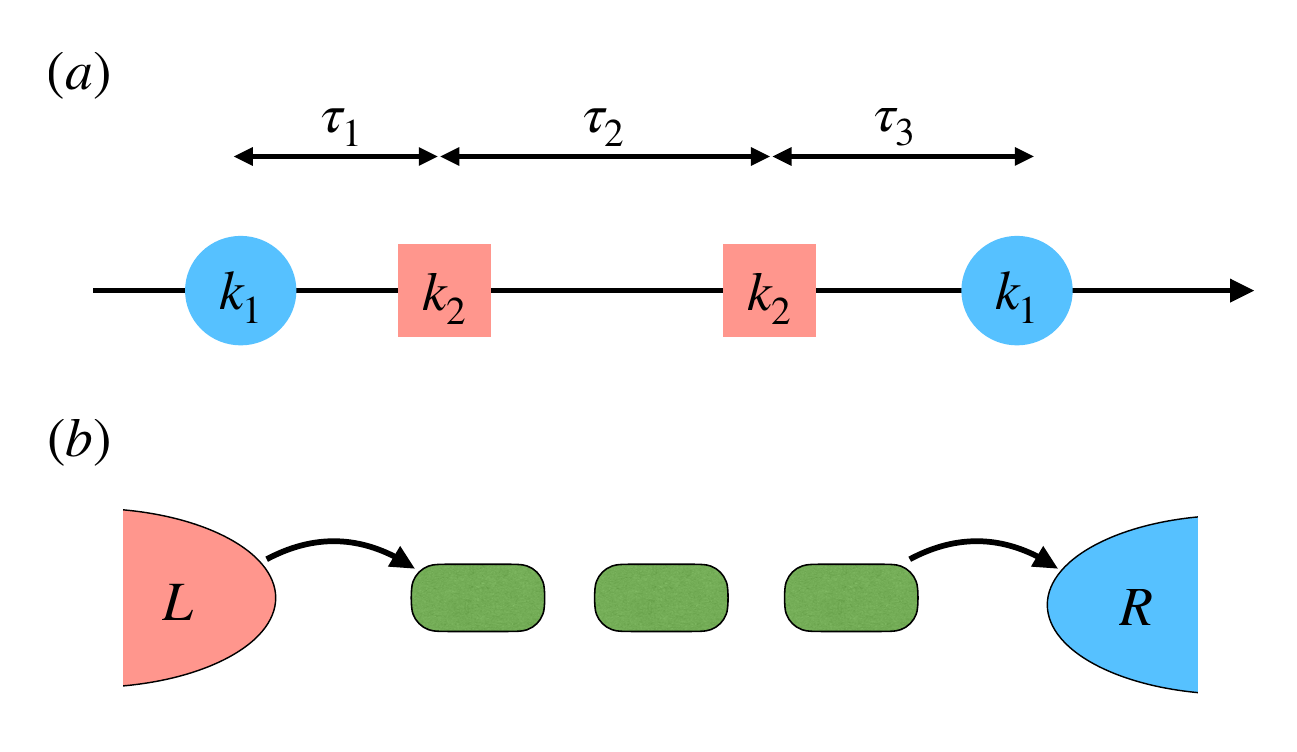}
    \caption{(a) A quantum jump trajectory viewed as a sequence of emissions, occurring at random channels and random times.
    Each jump channel $k_i$ can take one of two values in this case, here denoted by different colors.
    (b) Example of a system where multi-channel statistics appears: A boundary driven quantum chain connected to two reservoirs at each end. There are 2 jump channels, corresponding to injection ($I$) on the left  and emission ($E$) on the right.
    The alphabet of monitored channels is therefore $\mathbb{M} = \{I,E\}$.
    }
    \label{fig:drawing}
\end{figure}

\emph{Introduction - } We cannot see quantum systems. All we can do is perform measurements and analyze the resulting random outcomes.
In continuously measured systems~\cite{wiseman2009,jacobs2014,landi2023} these come in the form of a classical stochastic time series.
More specifically, in the jump unravelling~\cite{cook1985,plenio1998a,carmichael2014,nagourney1986a,sauter1986a,bergquist1986a,gustavsson2006} the dynamics is described by a series of abrupt jumps occurring at random intervals $\tau_i$ and at random channels $k_i$ (Fig.~\ref{fig:drawing}(a)):
\begin{equation}\label{trajectory}
    (\tau_1,k_1), (\tau_2,k_2),\ldots, (\tau_N,k_N).
\end{equation}
Each channel can be thought of as a  detector, which clicks when that jump occurs.
For instance, Fig.~\ref{fig:drawing}(b) shows a quantum chain with two channels,  corresponding to the injection ($I$) of an excitation on the left and extraction ($E$)  on the right. 
The outcomes $k_i$ therefore run over a certain alphabet of symbols, $\mathbb{M} = \{I,E\}$, labeling the different \emph{monitored channels}.
There has been a large number of studies dedicated to  the waiting times $\tau_i$ between two jumps~\cite{brandes2008,brandes2016,kosov2016,ptaszynski2017,walldorf2018,kleinherbers2021,carmichaelPhotoelectronWaitingTimes1989,vyas1988,albert2011,albert2012,thomas2013,rajabi2013,haack2014,thomas2014,dasenbrook2015}.
However,  little is known about the statistics of the jump channels $k_i$; that is, the distribution $\mathcal{P}(k_1,\ldots,k_N)$ that jump $k_1$ is followed by $k_2$ and so on.
This is relevant because quantum-coherent stochastic processes can have highly correlated outcomes with complex memory patterns~\cite{pollock2018a,pollock2018,milz2021,liu2019a,binder2018a,liu2018,yang2018}. 
Decoding these patterns therefore allow us to learn valuable information about the system.

Quantum systems with multiple jump channels have been extensively studied in the context of transport~\cite{landi2022a,bertini2021,prosen2008,benenti2009,karevski2009,prosen2010,dzhioev2011,popkov2012,mendoza-arenas2013a,mendoza-arenas2013}, dissipation-driven phase transitions~\cite{novotny2003,flindt2004a,minganti2018,guo2017}, current rectification~\cite{landi2014,landi2015,thingna2013,balachandran2018,balachandran2019,mendoza-arenas2022}, disorder~\cite{rebentrost2009,chiaracane2021a,lacerda2021a,varma2019a}, among others. 
These studies focused on the average current between the reservoirs.
Going beyond that,  Full Counting Statistics (FCS)~\cite{landi2023,schaller2014,esposito2009,levitov1993,levitov1996,nazarov2003,flindt2010} provides the  distribution of the accumulated (stochastic) charge after a fixed final time. 
Conversely, to obtain the statistics of~\eqref{trajectory} one must work in an ensemble where the number of jumps is fixed instead.
Formal results in this direction were put forth in~\cite{budini2014,kiukas2015}, and an algorithm to efficiently simulate these trajectories was introduced in~\cite{radaelli2023}.
Notwithstanding, very little is understood on how the features of the quantum system are encoded in the resulting statistics. 
This could be relevant, for example, in furthering our understanding of  measurement-induced phase transitions~\cite{turkeshi2021a,skinner2019,li2018a,coppola2022,alberton2021,carollo2022,cao2019}.

In this letter we address  how to uncover the memory structure and the resulting patterns present in a sequence of observations $(k_1,\ldots, k_N)$ of the jump channels, and how these relate to the underlying quantum features.
We find that the multi-jump statistics $\mathcal{P}(k_1,\ldots,k_N)$ is determined by the spectral properties of certain superoperators. 
This allows us to simulate this statistics efficiently, and also determine the conditions for the process to be stationary. 
Using this, we show that some systems support \emph{patterns}, in which the dynamics is fully described by transitions between a well-defined set of quantum states, similar in spirit to $\epsilon$-machines~\cite{crutchfield1989,shalizi2001,shalizi2004,crutchfield2012} and $q$-simulators~\cite{gu2012,mahoney2016,yang2018,binder2018a}.
These patterns are not present in all systems however.
Some only have recurring states and others, not even that. 
Notwithstanding, we show that it is still possible to cluster based on their predictions about future outcomes. 
Our results are applied to a boundary driven quantum XY chain and reveal unique features of  coherent transport which are not visible in currents or in any FCS quantity.

\emph{Theory - } We consider a system with Hamiltonian $H$ evolving under a quantum master equation of the form 
\begin{equation}\label{M}
    \frac{d\rho}{dt} = \mathcal{L}\rho = -i[H,\rho] + \sum_{k=1}^\size L_k \rho L_k^\dagger - \frac{1}{2} \{ L_k^\dagger L_k,\rho\}. 
\end{equation}
where $L_k$ are arbitrary jump operators.
Throughout we assume the master equation has a single steady-state $\mathcal{L}\rhoss = 0$. 
We consider the unravelling of~\eqref{M} in terms of quantum jumps~\cite{cook1985,plenio1998a,carmichael2014,nagourney1986a,sauter1986a,bergquist1986a,gustavsson2006}, as governed by the superoperators 
$\mathcal{J}_k\rho  = L_k \rho L_k^\dagger$.
For generality, we allow for the possibility that only a subset $\mathbb{M}$ of the jumps can be monitored. 
We therefore define the no-jump superoperator as $\mathcal{L}_0 = \mathcal{L} - \sum_{k\in \mathbb{M}} \mathcal{J}_k$.
For any initial state $\rho$, the probability of observing the trajectory~\eqref{trajectory} is given by~\cite{plenio1998a,budini2014,kiukas2015,landi2023}:
\begin{equation}\label{P_general}
    \mathcal{P}(\tau_1,k_1,\ldots,\tau_N,k_N) = \tr\big\{ 
    \mathcal{J}_{k_N} e^{\mathcal{L}_0 \tau_N} \ldots \mathcal{J}_{k_1} e^{\mathcal{L}_0 \tau_1} \rho \big\},
\end{equation}
with $k_i \in \mathbb{M}$ and $\tau_i \geqslant0$.
Marginalizing over the $k_i$ gives the joint probability of the waiting times $\tau_i$ alone. 
Instead, our  interest here will be in the statistics of jump channels $k_i$, obtained by integrating Eq.~\eqref{P_general} over all $\tau_i \geqslant 0$: 
\begin{equation}\label{P_channels}
    \mathcal{P}(k_1,\ldots,k_N) = \tr\big\{ \mathcal{M}_{k_N} \ldots \mathcal{M}_{k_1} \rho \big\},
\end{equation}
where $\mathcal{M}_k = - \mathcal{J}_k \mathcal{L}_0^{-1}$.
The existence of $\mathcal{L}_0^{-1}$ is tantamount to the hypothesis that there are no dark subspaces; i.e., that a jump must always eventually occur. We will assume this is the case.
Eq.~\eqref{P_channels} gives the probability of observing specific sequences (or \emph{patterns}) of clicks within the monitored set $\mathbb{M}$ (Fig.~\ref{fig:drawing}(a)).
We see that this is entirely governed by the superoperators $\mathcal{M}_k$. 

\emph{Stationarity and memory - }Marginalizing Eq.~\eqref{P_general} over any pair $(t,k)$ amounts to the replacement 
\begin{equation}\label{replacement_M}
\sum_{k\in \mathbb{M}} \int\limits_0^\infty dt \mathcal{J}_k e^{\mathcal{L}_0 t} = \mathcal{M} := \sum_{k\in \mathbb{M}} \mathcal{M}_k .
\end{equation}
As we will see, this superoperator $\mathcal{M}$ ultimately determines the memory structure of $\mathcal{P}(k_1,\ldots,k_N)$. 
One may verify that it satisfies $\tr\big\{ \mathcal{M}(\bullet)\big\} = \tr\big\{\bullet \big\}$, which follows from the fact that $\tr\big\{\mathcal{L}(\bullet)\big\} = 0$ for any Liouvillian. 
This property ensures that Eqs.~\eqref{P_general} and~\eqref{P_channels} are properly normalized. 

A stochastic process is stationary when $\mathcal{P}(k_1,\ldots,k_N) = \mathcal{P}(k_{i+1},\ldots,k_{i+N})$ for any $i,N$. 
With a generic initial state $\rho$, Eqs.~\eqref{P_general} or~\eqref{P_channels} are not stationary. 
As shown in~\cite{landi2023,zotero-1447}, this will only be the case if the initial state is 
\begin{equation}\label{pi}
    \pi = \frac{\mathcal{J}\rhoss}{K},
\end{equation}
which we call the \emph{Jump Steady-State} (JSS). 
Here $\mathcal{J} = \sum_{k\in\mathbb{M}} \mathcal{J}_k$ and $K = \tr(\mathcal{J}\rho)$ is the dynamical activity(number of jumps per unit time~\cite{landi2023}). 
As far as the jump dynamics is concerned, the relevant state is therefore not the steady-state $\rhoss$, but rather the JSS $\pi$ in Eq.~\eqref{pi}.
Henceforth, we assume the initial state is always $\rho = \pi$. 
In addition to stationarity, the state $\pi$ also yields other features. 
For example, since $\mathcal{L} = \mathcal{J} + \mathcal{L}_0$ and $\mathcal{L}\rhoss = 0$, it follows that in the JSS the single-outcome distribution is
\begin{equation}\label{Pk1}
    \mathcal{P}(k_1) = \tr\big\{ \mathcal{M}_{k_1}\pi\big\} =  \frac{1}{K}\tr\big\{ \mathcal{J}_{k_1} \rhoss\big\},
\end{equation}
which is  the relative frequency with which jump $k_1$ occurs in the steady-state. 


The JSS satisfies $\mathcal{M}\pi = \pi$; i.e., it is the right eigenvector of $\mathcal{M}$ with eigenvalue 1.
In~\cite{zotero-1447} we show that the other eigenvalues satisfy $|\mu_j|\leqslant 1$. 
Marginalizing Eq.~\eqref{P_channels} over $k_2,\ldots,k_{N-1}$ yields 
\begin{equation}\label{Pk1kN}
    \mathcal{P}(k_1,k_N) = \tr\big\{ \mathcal{M}_{k_N} \mathcal{M}^{N-2} \mathcal{M}_{k_1} \pi \big\},
\end{equation}
which, as shown in~\cite{zotero-1447}, can also be written as 
\begin{equation}\label{Pk1kN_spectral}
    \mathcal{P}(k_1,k_N) = \mathcal{P}(k_1)\mathcal{P}(k_N) + \sum_j \mu_j^{N-2} \tr\big\{ \mathcal{M}_{k_N} \mathbb{P}_j \mathcal{M}_{k_1} \pi \big\},
\end{equation}
where $\mathbb{P}_j$ are the eigenprojectors of $\mathcal{M}$ with eigenvalue $\mu_j$. 
We therefore see that the memory structure of $\mathcal{P}(k_1,k_N)$ is entirely determined by the spectrum of $\mathcal{M}$. 
This kind of memory structure can now be systematically analyzed using information-theoretic tools~\cite{wyner1978,crutchfield1989,renner2002,cover1991,yang2020,jurgens2021}. For example, the mutual information between $k_1$ and $k_N$ reads $\mathcal{I}(k_1\!\! : \!\! k_N) = \sum_{k_1,k_N} \mathcal{P}(k_1,k_N) \ln \mathcal{P}(k_1,k_N)/\mathcal{P}(k_1)\mathcal{P}(k_N)$.

\emph{Stochastic dynamics - }From Eq.~\eqref{P_channels}, the probability that a future outcome $k_{N+1}$ is observed, given a sequence of observations $k_1,\ldots,k_N$ may also be written as 
\begin{equation}\label{P_conditional}
    \mathcal{P}(k_{N+1} | k_1,\ldots, k_N) = \tr\big\{ \mathcal{M}_{k_{N+1}} \rho_{k_1\ldots k_N}\big\},
\end{equation}
where
$\rho_{k_1\ldots k_N} =  \mathcal{M}_{k_N} \ldots \mathcal{M}_{k_1} \rho/\tr\big\{ \mathcal{M}_{k_N} \ldots \mathcal{M}_{k_1} \rho \big\}.$
This  introduces a discrete-step stochastic dynamics, where after each jump the system evolves as
\begin{equation}\label{rho_dynamics}
    \rho_{i+1} = \frac{\mathcal{M}_{k_{i+1}} \rho_i}{\tr(\mathcal{M}_{k_{i+1}} \rho_i)},
\end{equation}
in which $k_{i+1}\in \mathbb{M}$ is sampled with probability $\tr\{\mathcal{M}_{k_{i+1}} \rho_i\big\}$.
The states $\rho_i$ are not associated to specific times.
They are the states of the system immediately after a jump when one is ignorant about the time between jumps. 
This is why the relevant superoperators are $\mathcal{M}_k = -\mathcal{J}_k \mathcal{L}_0^{-1}$ and not $\mathcal{J}_k$. 
The process in question is  a quantum version of a hidden Markov model. 
The system dynamics (the ``hidden layer'') evolves according to Eq.~\eqref{rho_dynamics}, which is a (non-linear) completely positive trace preserving map (as we prove in~\cite{zotero-1447}).
But what is actually observed (the ``visible layer'') is a sequence of emitted symbols: each transition $\rho_i \to \rho_{i+1}$ is associated with a unique symbol $k_{i+1}\in \mathbb{M}$.
Processes of this form are called unifilar~\cite{crutchfield1989,shalizi2001,shalizi2004,crutchfield2012,travers2012}.

\begin{figure}
    \centering
    \includegraphics[width=0.45\textwidth]{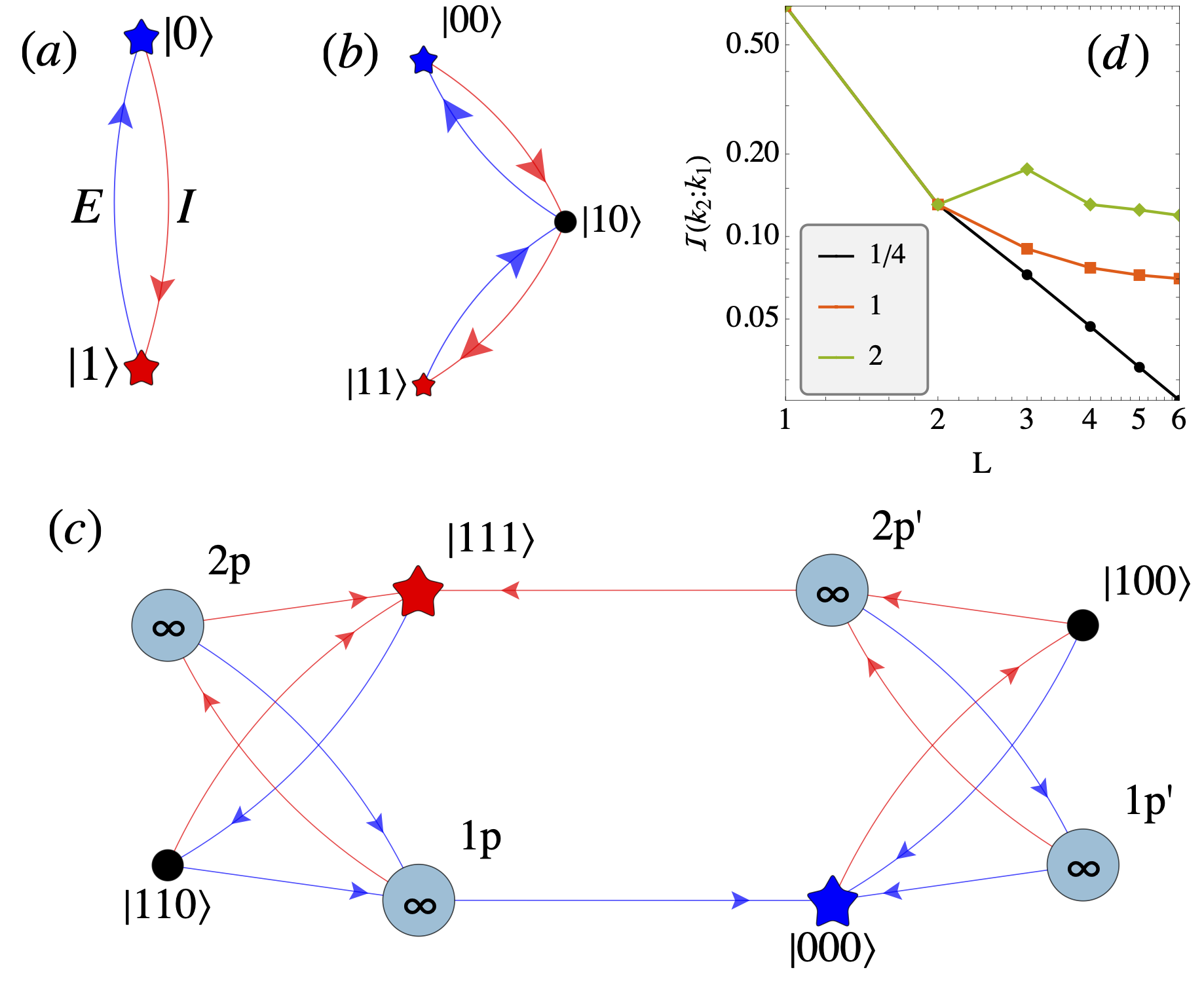}
    \caption{(a)-(c) Patterns emerging in the boundary-driven XX chain, for $L = 1,2,3$ respectively.
    Red arrows are injection jumps, $\mathcal{M}_I$, while blue arrows refer to extraction, $\mathcal{M}_E$. 
    (a) $L=1$: the process is renewal;
    (b) $L=2$: process is not renewal, but still supports a closed pattern. 
    (c) $L=3$: no closed pattern, but recurring states exist. 
    (e) Log-log plot of the mutual information between two outcomes $\mathcal{I}(k_1\!\! : \!\! k_2)$ as a function of $L$ for different values of $\gamma/L$.
    }
    \label{fig:patterns}
\end{figure}

\emph{Patterns - }To uncover the structure and features of the multi-jump probability~\eqref{P_channels} we must understand the states that are sampled as the system evolves according to Eq.~\eqref{rho_dynamics}.
More precisely, one must ask whether there are any \emph{patterns}, or any \emph{predictability}, to this evolution. 
For example, do the states ever repeat? 
Patterns are relevant since they determine the memory about the past that must be stored in order to predict the future, as per Eq.~\eqref{P_conditional}.
We define a \emph{closed pattern} as a finite set of states $\mathbb{B} = \{\sigma_1,\sigma_2,\ldots,\}$  for which the dynamics under Eq.~\eqref{rho_dynamics} is closed; i.e., such that 
\begin{equation}\label{pattern}
    \frac{\mathcal{M}_k\sigma_i}{\tr(\mathcal{M}_k\sigma_i)} = \sigma_{f(i,k)} \in \mathbb{B},
\end{equation}
for a function $f\!\!:\!\! (\mathbb{B},\mathbb{M}) \to \mathbb{B}$.
The $\sigma_i$ are the analog of the causal states in computational mechanics~\cite{crutchfield1989,shalizi2001,shalizi2004,crutchfield2012}.
If the system is ever in a closed pattern, it remains in it throughout the dynamics.
We may therefore represent a pattern by a graph, with nodes labeled by $\sigma_i \in \mathbb{B}$ and edges labeled by the jump channels $k \in \mathbb{M}$. 
For each node $i$ there are at most $|\mathbb{M}|$ outgoing edges (the number can be smaller than $|\mathbb{M}|$ because for some states $\mathcal{M}_k \sigma_i = 0$). 
Examples of pattern graphs are shown in Fig.~\ref{fig:patterns}(a) and (b).

A class of processes that support closed patterns are the renewal process~\cite{brandes2008}, in which a jump completely resets the state of the system;  
that is, $\mathcal{M}_k \rho/\tr(\mathcal{M}_k \rho) = \sigma_k$, $\forall \rho$.
The size of the pattern is thus $|\mathbb{B}| = |\mathbb{M}|$.
Nodes and edges are labeled by the same alphabet. 
And edge $k$ always points to node $k$ (c.f Fig.~\ref{fig:patterns}(a)). 
Renewal processes appear in classical (Pauli) master equations~\cite{vankampen2007} or
in Global master equations (Davies maps) when the transition frequencies are non-degenerate~\cite{breuer2007,davies1974,gonzalez2017,wichterich2007}. 
In the latter, the $\sigma_k$ are the energy eigenstates. 
For renewal processes  Eq.~\eqref{P_conditional} reduces to 
$\mathcal{P}(k_{N+1}|k_1,\ldots,k_N) = \tr\big\{ \mathcal{M}_{k_{N+1}} \sigma_{k_N}\big\} \equiv \mathcal{P}(k_{N+1}|k_N)$,
which is a Markov order 1 process. 

Not all systems support closed patterns, however. 
A weaker form of predictability is when the process has a \emph{recurring state}. 
That is, a state $\sigma_0$ such that, starting from an arbitrary initial state $\rho$, Eq.~\eqref{rho_dynamics} will always eventually pass through $\sigma_0$ with probability 1. 
This is reminiscent of  Feller's theory of recurring events~\cite{feller1949}. 
Depending on the type of jump operators and Hamiltonians, however, some processes might not even have recurring states, as we exemplify below. 


Closed patterns and recurring states are relevant because they make it easier to  predict future outcomes. 
To predict $k_{N+1}$ we need to use Eq.~\eqref{P_conditional}, which requires all outcomes $k_1,\ldots,k_N$.
But if we know in which pattern state the system is in, that is not necessary. 
The existence of a closed pattern define a minimal sufficient statistic for predicting future outcomes~\cite{shalizi2004}. 
It means we can bundle different histories together whenever they predict the same future outcomes. 
That is, if $\mathcal{P}(k_{N+1}|k_1\ldots k_N)=\mathcal{P}(k_{N+1}|k_1'\ldots k_N')$, it is because $k_1\ldots k_N$ and $k_1'\ldots k_N'$ actually correspond to the same pattern state $\sigma_i$, as is evident from Eq.~\eqref{P_conditional}.

\emph{Boundary driven spin chain - }We exemplify our results with a one-dimensional qubit (spin 1/2) chain with $L$ sites, connected to two reservoirs at each boundary~\cite{landi2022a}. 
The Hamiltonian is taken as the XX model 
$H = J \sum_{i=1}^{L-1} \big( \sigma_+^i \sigma_-^{i+1} + \sigma_-^i \sigma_+^{i+1}\big)$,
where $J$ is the overall energy scale.
The master equation is chosen to mimic the scenario in Fig.~\ref{fig:drawing}(b):
$\mathcal{L}\rho = -i[H,\rho] + \gamma \mathcal{D}[\sigma_1^+]\rho + \gamma \mathcal{D}[\sigma_L^-]\rho$,
so that $\mathcal{J}_I \rho = \gamma \sigma_1^+ \rho \sigma_1^-$ describes the injection ($I$) of excitations on site 1 and 
$\mathcal{J}_E \rho = \gamma \sigma_L^- \rho \sigma_L^+$ describes 
the extraction ($E$) on site $L$. 
In the steady-state, the current of excitations leaving the system is 
$J = \gamma \langle \sigma_L^+ \sigma_L^-\rangle = -\gamma \langle \sigma_1^+ \sigma_1^-\rangle$.
The current is thus related to Eq.~\eqref{Pk1} according to 
$J = \gamma \mathcal{P}(E)$.

Fig.~\ref{fig:patterns} shows the patterns obtained when $L = 1,2,3$ (details on how they are determined are given in~\cite{zotero-1447}).
The case $L=1$ is renewal (Fig.~\ref{fig:patterns}(a)). 
The pattern has two states $|0\rangle, |1\rangle$ and transitions deterministically as
$|1\rangle \xrightarrow[]{E} |0\rangle$ and 
$|0\rangle \xrightarrow[]{I} |1\rangle$. 
Conversely, $L=2$ is not renewal because if a jump happens in qubit 1, it will not completely reset the state of all qubits together. 
Notwithstanding, the system still admits a closed pattern
(Fig.~\ref{fig:patterns}(b)): by loosing excitations the system goes from $|11\rangle\xrightarrow[]{E} |10\rangle \xrightarrow[]{E} |00\rangle$, and the other way around with $\mathcal{M}_I$.
From $|11\rangle$ and $|00\rangle$ the transitions are deterministic, but from $|10\rangle$ they are not (there are two arrows coming out of it).

For $L\geqslant 3$ the situation becomes much more complicated: the system no longer admits a closed pattern, although it still has recurring states. 
The case $L=3$ is shown in Fig.~\ref{fig:patterns}(c).
From $|111\rangle$ the system goes deterministically to $|110\rangle$ via $E$.
From $|110\rangle$ it can either go back to $|111\rangle$ via $I$, or go to a new state which we call $\sigma_{\rm 1p}^{(1)}$, which is \emph{not} $|100\rangle$:
it has exactly one particle, but is a mixed state, with non-zero off-diagonals (coherences).
From $\sigma_{\rm 1p}^{(1)}$
an $E$ jump would take the system to $|000\rangle$ and hence reset the process. 
But an $I$ jump  takes it to a 2-particle state $\sigma_{\rm 2p}^{(2)}$, which is also mixed and coherent. 
Finally, from $\sigma_{\rm 2p}^{(2)}$ the state can be reset to $|111\rangle$ via $I$ or it can go to yet another 1-particle state $\sigma_{\rm 1p}^{(3)}$ which is different from $\sigma_{\rm 1p}^{(1)}$ (see~\cite{zotero-1447} for more details).
From $|000\rangle$ the system would sample a similar trajectory, but through sets 1p$'$ and 2p$'$ which are different from 1p and 2p.
As long as the system keeps  bouncing back and forth between 1p and 2p (or 1p$'$ and 2p$'$) it will continue to explore new pattern states that never repeat. 
The sampled states are therefore infinite, although they are all recurring, some more than others.  
Because of the existence of these recurring states, particularly, $|111\rangle$ and $|000\rangle$, there is still a significant truncation in the memory.

Fig.~\ref{fig:patterns}(d)  shows the mutual information between two outcomes, 
$\mathcal{I}(k_1\!\! : \!\! k_2)$ as a function of $L$ for different choices of $\gamma/L$. 
For $L = 1,2$ the correlations are independent of the ratio $\gamma/L$, which is no longer true for $L\geqslant 3$.
The results indicate that in the thermodynamic limit the outcomes become uncorrelated, which is expected. 
For instance, an emission becomes less dependent on whether the previous jump was an emission or an injection, since there are already several excitations inside the chain.

\emph{Pairing terms - }The  process becomes dramatically more complex if we also include  pairing terms $H_{\rm pair} = \kappa \sum_{i=1}^{L-1} \big( \sigma_i^+ \sigma_{i+1}^+ + \sigma_i^- \sigma_{i+1}^-\big)$  (i.e., turn the XX chain into an XY chain). 
In between jumps, the number of particles can  fluctuate arbitrarily, although always in pairs. 
As a consequence, the dynamics has no recurring states~\cite{zotero-1447}.
In this case, we can try to determine \emph{approximate} closed patterns. 
Following~\cite{crutchfield1989,shalizi2001,shalizi2004,crutchfield2012}, we do this by grouping states that produce the same future outcomes.
That is, two states $\rho_1$ and $\rho_2$ belong to the same cluster if $\mathcal{P}(k_1,k_2,\ldots,|\rho_1)$ and $\mathcal{P}(k_1,k_2,\ldots,|\rho_2)$ are  close up to some tolerance.
We apply the following methodology. 
First we run Eq.~\eqref{rho_dynamics} for a long time and discard the first steps to eliminate transients.
The remaining states are then clustered according to their probabilities in predicting the future, $\mathcal{P}(k_1,\ldots,k_n|\rho) = \tr\big\{ \mathcal{M}_{k_n} \ldots \mathcal{M}_{k_1} \rho)$ for some fixed $n$ (we choose $n=6$). 
We employ hierarchical agglomerative clustering with single-linkage and distance function  
$D(\rho_1,\rho_2)^2 = \sum_{k_1,\ldots,k_n} \big[\mathcal{P}(k_1,\ldots,k_n|\rho_1)-\mathcal{P}(k_1,\ldots,k_n|\rho_2)\big]^2$~\footnote{
We could also have used as a distance function the trace-distance between the two density matrices. 
However, we believe that clustering based on probabilities is more meaningful as it will bundle together states which differ not in their shapes, but in how they affect the statistics.}. 

\begin{figure}
    \centering
    \includegraphics[width=0.5\textwidth]{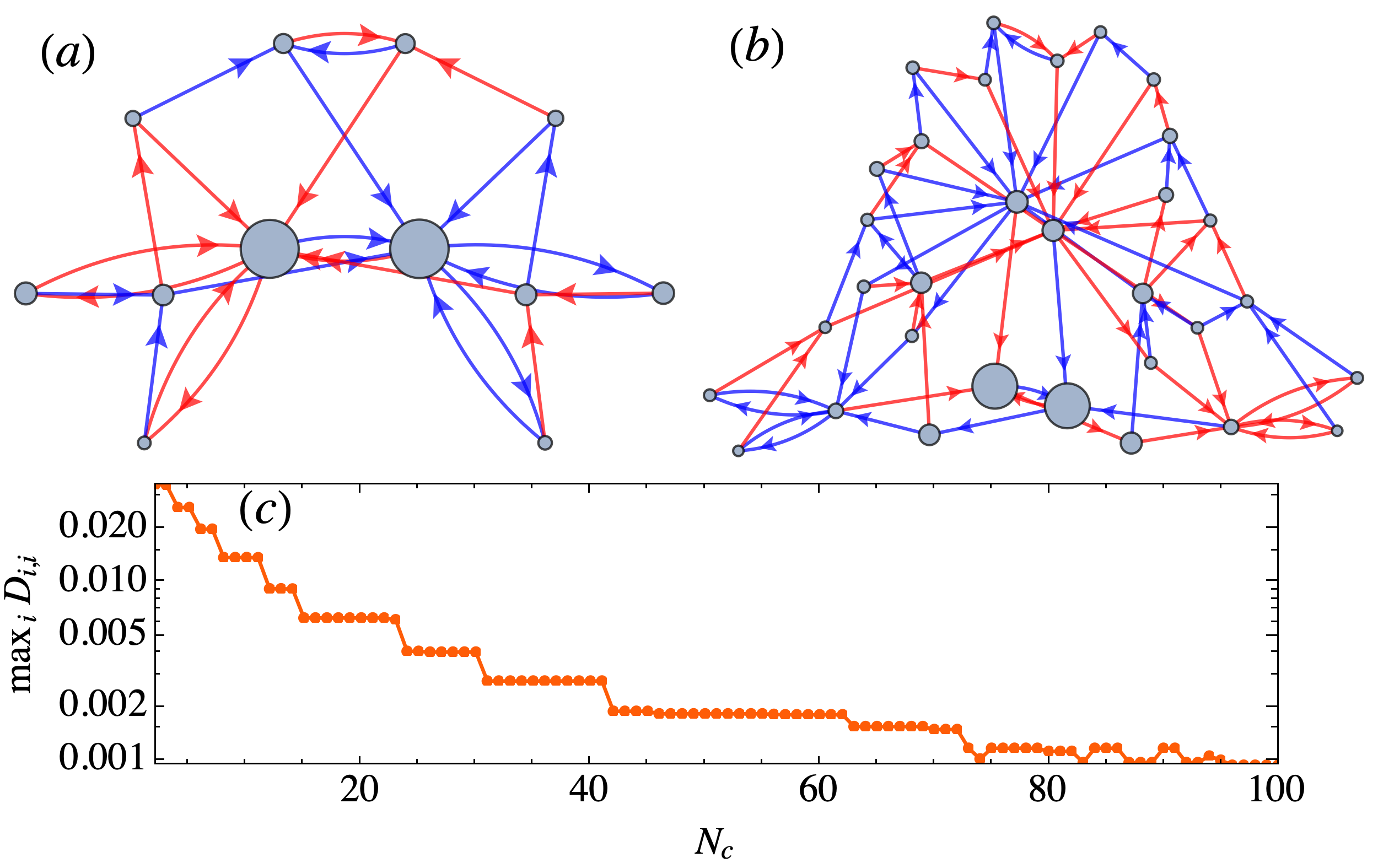}
    \caption{Pattern graphs for the $XY$ model with $L=3$, $\gamma/J = 1$ and $\kappa/J = 1/2$. 
    (a) With $12$ clusters.
    (b) With $32$ clusters.
    The size of the nodes, as well as the number of outgoing arrows, is proportional to the relative number of states within each cluster. 
    (c) Maximum distance between states within each cluster, as a function of the number of clusters $N_c$. 
    }
    \label{fig:kappa_graphs}
\end{figure}

We draw pattern graphs for different number of clusters $N_c$. 
Examples with $N_c = 12$ and $32$ are shown in Fig.~\ref{fig:kappa_graphs}(a),(b).
The size of the nodes reflects the number of states in each cluster. 
For each $N_c$ we also build a distance matrix between clusters by averaging the distances between each state in each cluster. 
That is, if $S_i$ is the set of states belonging to cluster $i$ (with $i = 1,\ldots,N_c$) then 
$D_{i,j} = \frac{1}{|S_i| |S_j|} \sum_{k \in S_i, q\in S_j} D(\rho_k,\rho_q)$.
The diagonal entries $D_{i,i}$  measure  the typical distances between states within the same cluster. 
In Fig.~\ref{fig:kappa_graphs}(c) we plot $\max_i D_{i,i}$ as a function of $N_c$. 
This gives a measure of the overall quality of the clustering. 
To perfectly predict the future we need an infinite number of clusters, as anticipated.
However, the quality improves gradually with $N_c$.

\emph{Discussion -}We showed how to obtain the full multi-point statistics of jump channels in continuously monitored systems. Processes involving multiple jump channels are quite common.
But their multi-time statistics has so far been largely unexplored. 
Here we showed that the jump-channel statistics is given by a discrete-space and discrete-step stochastic process with intricate memory patterns, reflecting the features of the underlying quantum dynamics. 
We also provided a comprehensive framework to characterize and understand them.
Eq.~\eqref{P_channels} contains a richness of information that goes far beyond the average current or FCS.
This is relevant because patterns in the string of jumps is what is  experimentally observed~\cite{cook1985,plenio1998a,nagourney1986a,sauter1986a,bergquist1986a,gustavsson2006,fink2018,fink2017}. 
And we can use these to learn about the quantum system. 
Consider, for example, a string such as $EIIEEIEIEEIIEIE$.
What is the likelihood that this was generated by a chain of length $L=3$?
And what is the likelihood that pairing terms are present?
With these tools, this question can now be easily addressed. 

\emph{Acknowledgments - }The author acknowledges fruitful discussions with M. Kewming, G. Ghoshal, J. Smiga, A. Hedge and G. Chandrasekharan.

\bibliography{library}

\begin{thebibliography}{94}%
\makeatletter
\providecommand \@ifxundefined [1]{%
 \@ifx{#1\undefined}
}%
\providecommand \@ifnum [1]{%
 \ifnum #1\expandafter \@firstoftwo
 \else \expandafter \@secondoftwo
 \fi
}%
\providecommand \@ifx [1]{%
 \ifx #1\expandafter \@firstoftwo
 \else \expandafter \@secondoftwo
 \fi
}%
\providecommand \natexlab [1]{#1}%
\providecommand \enquote  [1]{``#1''}%
\providecommand \bibnamefont  [1]{#1}%
\providecommand \bibfnamefont [1]{#1}%
\providecommand \citenamefont [1]{#1}%
\providecommand \href@noop [0]{\@secondoftwo}%
\providecommand \href [0]{\begingroup \@sanitize@url \@href}%
\providecommand \@href[1]{\@@startlink{#1}\@@href}%
\providecommand \@@href[1]{\endgroup#1\@@endlink}%
\providecommand \@sanitize@url [0]{\catcode `\\12\catcode `\$12\catcode
  `\&12\catcode `\#12\catcode `\^12\catcode `\_12\catcode `\%12\relax}%
\providecommand \@@startlink[1]{}%
\providecommand \@@endlink[0]{}%
\providecommand \url  [0]{\begingroup\@sanitize@url \@url }%
\providecommand \@url [1]{\endgroup\@href {#1}{\urlprefix }}%
\providecommand \urlprefix  [0]{URL }%
\providecommand \Eprint [0]{\href }%
\providecommand \doibase [0]{http://dx.doi.org/}%
\providecommand \selectlanguage [0]{\@gobble}%
\providecommand \bibinfo  [0]{\@secondoftwo}%
\providecommand \bibfield  [0]{\@secondoftwo}%
\providecommand \translation [1]{[#1]}%
\providecommand \BibitemOpen [0]{}%
\providecommand \bibitemStop [0]{}%
\providecommand \bibitemNoStop [0]{.\EOS\space}%
\providecommand \EOS [0]{\spacefactor3000\relax}%
\providecommand \BibitemShut  [1]{\csname bibitem#1\endcsname}%
\let\auto@bib@innerbib\@empty
\bibitem [{\citenamefont {Wiseman}\ and\ \citenamefont
  {Milburn}(2009)}]{wiseman2009}%
  \BibitemOpen
  \bibfield  {author} {\bibinfo {author} {\bibfnamefont {H.~M.}\ \bibnamefont
  {Wiseman}}\ and\ \bibinfo {author} {\bibfnamefont {G.~J.}\ \bibnamefont
  {Milburn}},\ }\href@noop {} {\emph {\bibinfo {title} {Quantum measurement and
  control}}}\ (\bibinfo  {publisher} {Cambridge University Press},\ \bibinfo
  {address} {New York},\ \bibinfo {year} {2009})\BibitemShut {NoStop}%
\bibitem [{\citenamefont {Jacobs}(2014)}]{jacobs2014}%
  \BibitemOpen
  \bibfield  {author} {\bibinfo {author} {\bibfnamefont {K.}~\bibnamefont
  {Jacobs}},\ }\href@noop {} {\emph {\bibinfo {title} {Quantum measurement
  theory and its applications}}}\ (\bibinfo  {publisher} {Cambridge University
  Press},\ \bibinfo {address} {Cambridge},\ \bibinfo {year} {2014})\BibitemShut
  {NoStop}%
\bibitem [{\citenamefont {Landi}\ \emph {et~al.}(2023)\citenamefont {Landi},
  \citenamefont {Kewming}, \citenamefont {Mitchison},\ and\ \citenamefont
  {Potts}}]{landi2023}%
  \BibitemOpen
  \bibfield  {author} {\bibinfo {author} {\bibfnamefont {G.~T.}\ \bibnamefont
  {Landi}}, \bibinfo {author} {\bibfnamefont {M.~J.}\ \bibnamefont {Kewming}},
  \bibinfo {author} {\bibfnamefont {M.~T.}\ \bibnamefont {Mitchison}}, \ and\
  \bibinfo {author} {\bibfnamefont {P.~P.}\ \bibnamefont {Potts}},\ }\href
  {http://arxiv.org/abs/2303.04270} {\enquote {\bibinfo {title} {Current
  fluctuations in open quantum systems: {Bridging} the gap between quantum
  continuous measurements and full counting statistics},}\ } (\bibinfo {year}
  {2023}),\ \bibinfo {note} {arXiv:2303.04270 [cond-mat,
  physics:quant-ph]}\BibitemShut {NoStop}%
\bibitem [{\citenamefont {Cook}\ and\ \citenamefont {Kimble}(1985)}]{cook1985}%
  \BibitemOpen
  \bibfield  {author} {\bibinfo {author} {\bibfnamefont {R.~J.}\ \bibnamefont
  {Cook}}\ and\ \bibinfo {author} {\bibfnamefont {H.~J.}\ \bibnamefont
  {Kimble}},\ }\href {\doibase 10.1103/PhysRevLett.54.1023} {\bibfield
  {journal} {\bibinfo  {journal} {Physical Review Letters}\ }\textbf {\bibinfo
  {volume} {54}},\ \bibinfo {pages} {1023} (\bibinfo {year}
  {1985})}\BibitemShut {NoStop}%
\bibitem [{\citenamefont {Plenio}\ and\ \citenamefont
  {Knight}(1998)}]{plenio1998a}%
  \BibitemOpen
  \bibfield  {author} {\bibinfo {author} {\bibfnamefont {M.~B.}\ \bibnamefont
  {Plenio}}\ and\ \bibinfo {author} {\bibfnamefont {P.~L.}\ \bibnamefont
  {Knight}},\ }\href {\doibase 10.1103/RevModPhys.70.101} {\bibfield  {journal}
  {\bibinfo  {journal} {Reviews of Modern Physics}\ }\textbf {\bibinfo {volume}
  {70}},\ \bibinfo {pages} {101} (\bibinfo {year} {1998})},\ \bibinfo {note}
  {arXiv: quant-ph/9702007 ISBN: 0034-6861}\BibitemShut {NoStop}%
\bibitem [{\citenamefont {Carmichael}(2014)}]{carmichael2014}%
  \BibitemOpen
  \bibfield  {author} {\bibinfo {author} {\bibfnamefont {H.}~\bibnamefont
  {Carmichael}},\ }\href@noop {} {\emph {\bibinfo {title} {An {Open} {Systems}
  {Approach} to {Quantum} {Optics}}}},\ \bibinfo {edition} {3rd}\ ed.\
  (\bibinfo  {publisher} {Springer},\ \bibinfo {year} {2014})\BibitemShut
  {NoStop}%
\bibitem [{\citenamefont {Nagourney}\ \emph {et~al.}(1986)\citenamefont
  {Nagourney}, \citenamefont {Sandberg},\ and\ \citenamefont
  {Dehmelt}}]{nagourney1986a}%
  \BibitemOpen
  \bibfield  {author} {\bibinfo {author} {\bibfnamefont {W.}~\bibnamefont
  {Nagourney}}, \bibinfo {author} {\bibfnamefont {J.}~\bibnamefont {Sandberg}},
  \ and\ \bibinfo {author} {\bibfnamefont {H.}~\bibnamefont {Dehmelt}},\ }\href
  {\doibase 10.1103/PhysRevLett.56.2797} {\bibfield  {journal} {\bibinfo
  {journal} {Physical Review Letters}\ }\textbf {\bibinfo {volume} {56}},\
  \bibinfo {pages} {2797} (\bibinfo {year} {1986})}\BibitemShut {NoStop}%
\bibitem [{\citenamefont {Sauter}\ \emph {et~al.}(1986)\citenamefont {Sauter},
  \citenamefont {Neuhauser}, \citenamefont {Blatt},\ and\ \citenamefont
  {Toschek}}]{sauter1986a}%
  \BibitemOpen
  \bibfield  {author} {\bibinfo {author} {\bibfnamefont {T.}~\bibnamefont
  {Sauter}}, \bibinfo {author} {\bibfnamefont {W.}~\bibnamefont {Neuhauser}},
  \bibinfo {author} {\bibfnamefont {R.}~\bibnamefont {Blatt}}, \ and\ \bibinfo
  {author} {\bibfnamefont {P.~E.}\ \bibnamefont {Toschek}},\ }\href {\doibase
  10.1103/PhysRevLett.57.1696} {\bibfield  {journal} {\bibinfo  {journal}
  {Physical Review Letters}\ }\textbf {\bibinfo {volume} {57}},\ \bibinfo
  {pages} {1696} (\bibinfo {year} {1986})},\ \bibinfo {note} {publisher:
  American Physical Society}\BibitemShut {NoStop}%
\bibitem [{\citenamefont {Bergquist}\ \emph {et~al.}(1986)\citenamefont
  {Bergquist}, \citenamefont {Hulet}, \citenamefont {Itano},\ and\
  \citenamefont {Wineland}}]{bergquist1986a}%
  \BibitemOpen
  \bibfield  {author} {\bibinfo {author} {\bibfnamefont {J.~C.}\ \bibnamefont
  {Bergquist}}, \bibinfo {author} {\bibfnamefont {R.~G.}\ \bibnamefont
  {Hulet}}, \bibinfo {author} {\bibfnamefont {W.~M.}\ \bibnamefont {Itano}}, \
  and\ \bibinfo {author} {\bibfnamefont {D.~J.}\ \bibnamefont {Wineland}},\
  }\href {\doibase 10.1103/PhysRevLett.57.1699} {\bibfield  {journal} {\bibinfo
   {journal} {Physical Review Letters}\ }\textbf {\bibinfo {volume} {57}},\
  \bibinfo {pages} {1699} (\bibinfo {year} {1986})},\ \bibinfo {note}
  {publisher: American Physical Society}\BibitemShut {NoStop}%
\bibitem [{\citenamefont {Gustavsson}\ \emph {et~al.}(2006)\citenamefont
  {Gustavsson}, \citenamefont {Leturcq}, \citenamefont {Simovič},
  \citenamefont {Schleser}, \citenamefont {Ihn}, \citenamefont {Studerus},
  \citenamefont {Ensslin}, \citenamefont {Driscoll},\ and\ \citenamefont
  {Gossard}}]{gustavsson2006}%
  \BibitemOpen
  \bibfield  {author} {\bibinfo {author} {\bibfnamefont {S.}~\bibnamefont
  {Gustavsson}}, \bibinfo {author} {\bibfnamefont {R.}~\bibnamefont {Leturcq}},
  \bibinfo {author} {\bibfnamefont {B.}~\bibnamefont {Simovič}}, \bibinfo
  {author} {\bibfnamefont {R.}~\bibnamefont {Schleser}}, \bibinfo {author}
  {\bibfnamefont {T.}~\bibnamefont {Ihn}}, \bibinfo {author} {\bibfnamefont
  {P.}~\bibnamefont {Studerus}}, \bibinfo {author} {\bibfnamefont
  {K.}~\bibnamefont {Ensslin}}, \bibinfo {author} {\bibfnamefont {D.~C.}\
  \bibnamefont {Driscoll}}, \ and\ \bibinfo {author} {\bibfnamefont {A.~C.}\
  \bibnamefont {Gossard}},\ }\href {\doibase 10.1103/PhysRevLett.96.076605}
  {\bibfield  {journal} {\bibinfo  {journal} {Physical Review Letters}\
  }\textbf {\bibinfo {volume} {96}},\ \bibinfo {pages} {076605} (\bibinfo
  {year} {2006})}\BibitemShut {NoStop}%
\bibitem [{\citenamefont {Brandes}(2008)}]{brandes2008}%
  \BibitemOpen
  \bibfield  {author} {\bibinfo {author} {\bibfnamefont {T.}~\bibnamefont
  {Brandes}},\ }\href {\doibase 10.1002/andp.200810306} {\bibfield  {journal}
  {\bibinfo  {journal} {Annalen der Physik (Leipzig)}\ }\textbf {\bibinfo
  {volume} {17}},\ \bibinfo {pages} {477} (\bibinfo {year} {2008})},\ \bibinfo
  {note} {arXiv: 0802.2233}\BibitemShut {NoStop}%
\bibitem [{\citenamefont {Brandes}\ and\ \citenamefont
  {Emary}(2016)}]{brandes2016}%
  \BibitemOpen
  \bibfield  {author} {\bibinfo {author} {\bibfnamefont {T.}~\bibnamefont
  {Brandes}}\ and\ \bibinfo {author} {\bibfnamefont {C.}~\bibnamefont
  {Emary}},\ }\href {\doibase 10.1103/PhysRevE.93.042103} {\bibfield  {journal}
  {\bibinfo  {journal} {Physical Review E}\ }\textbf {\bibinfo {volume} {93}},\
  \bibinfo {pages} {042103} (\bibinfo {year} {2016})},\ \bibinfo {note} {arXiv:
  1602.02975}\BibitemShut {NoStop}%
\bibitem [{\citenamefont {Kosov}(2016)}]{kosov2016}%
  \BibitemOpen
  \bibfield  {author} {\bibinfo {author} {\bibfnamefont {D.~S.}\ \bibnamefont
  {Kosov}},\ }\href {http://arxiv.org/abs/1605.02170} {\enquote {\bibinfo
  {title} {Distribution of waiting times between superoperator quantum jumps in
  {Lindblad} dynamics},}\ } (\bibinfo {year} {2016}),\ \bibinfo {note}
  {arXiv:1605.02170 [cond-mat, physics:quant-ph]}\BibitemShut {NoStop}%
\bibitem [{\citenamefont {Ptaszyński}(2017)}]{ptaszynski2017}%
  \BibitemOpen
  \bibfield  {author} {\bibinfo {author} {\bibfnamefont {K.}~\bibnamefont
  {Ptaszyński}},\ }\href {\doibase 10.1103/PhysRevB.96.035409} {\bibfield
  {journal} {\bibinfo  {journal} {Physical Review B}\ }\textbf {\bibinfo
  {volume} {96}},\ \bibinfo {pages} {035409} (\bibinfo {year} {2017})},\
  \bibinfo {note} {arXiv: 1707.00441v1}\BibitemShut {NoStop}%
\bibitem [{\citenamefont {Walldorf}\ \emph {et~al.}(2018)\citenamefont
  {Walldorf}, \citenamefont {Padurariu}, \citenamefont {Jauho},\ and\
  \citenamefont {Flindt}}]{walldorf2018}%
  \BibitemOpen
  \bibfield  {author} {\bibinfo {author} {\bibfnamefont {N.}~\bibnamefont
  {Walldorf}}, \bibinfo {author} {\bibfnamefont {C.}~\bibnamefont {Padurariu}},
  \bibinfo {author} {\bibfnamefont {A.-P.}\ \bibnamefont {Jauho}}, \ and\
  \bibinfo {author} {\bibfnamefont {C.}~\bibnamefont {Flindt}},\ }\href
  {\doibase 10.1103/PhysRevLett.120.087701} {\bibfield  {journal} {\bibinfo
  {journal} {Physical Review Letters}\ }\textbf {\bibinfo {volume} {120}},\
  \bibinfo {pages} {087701} (\bibinfo {year} {2018})},\ \bibinfo {note} {arXiv:
  1709.01335}\BibitemShut {NoStop}%
\bibitem [{\citenamefont {Kleinherbers}\ \emph {et~al.}(2021)\citenamefont
  {Kleinherbers}, \citenamefont {Stegmann},\ and\ \citenamefont
  {König}}]{kleinherbers2021}%
  \BibitemOpen
  \bibfield  {author} {\bibinfo {author} {\bibfnamefont {E.}~\bibnamefont
  {Kleinherbers}}, \bibinfo {author} {\bibfnamefont {P.}~\bibnamefont
  {Stegmann}}, \ and\ \bibinfo {author} {\bibfnamefont {J.}~\bibnamefont
  {König}},\ }\href {\doibase 10.1103/PhysRevB.104.165304} {\bibfield
  {journal} {\bibinfo  {journal} {Physical Review B}\ }\textbf {\bibinfo
  {volume} {104}},\ \bibinfo {pages} {165304} (\bibinfo {year} {2021})},\
  \bibinfo {note} {arXiv: 2107.10218}\BibitemShut {NoStop}%
\bibitem [{\citenamefont {Carmichael}\ \emph {et~al.}(1989)\citenamefont
  {Carmichael}, \citenamefont {Singh}, \citenamefont {Vyas},\ and\
  \citenamefont {Rice}}]{carmichaelPhotoelectronWaitingTimes1989}%
  \BibitemOpen
  \bibfield  {author} {\bibinfo {author} {\bibfnamefont {H.~J.}\ \bibnamefont
  {Carmichael}}, \bibinfo {author} {\bibfnamefont {S.}~\bibnamefont {Singh}},
  \bibinfo {author} {\bibfnamefont {R.}~\bibnamefont {Vyas}}, \ and\ \bibinfo
  {author} {\bibfnamefont {P.~R.}\ \bibnamefont {Rice}},\ }\href {\doibase
  10.1103/PhysRevA.39.1200} {\bibfield  {journal} {\bibinfo  {journal}
  {Physical Review A}\ }\textbf {\bibinfo {volume} {39}},\ \bibinfo {pages}
  {1200} (\bibinfo {year} {1989})},\ \bibinfo {note} {publisher: American
  Physical Society}\BibitemShut {NoStop}%
\bibitem [{\citenamefont {Vyas}\ and\ \citenamefont {Singh}(1988)}]{vyas1988}%
  \BibitemOpen
  \bibfield  {author} {\bibinfo {author} {\bibfnamefont {R.}~\bibnamefont
  {Vyas}}\ and\ \bibinfo {author} {\bibfnamefont {S.}~\bibnamefont {Singh}},\
  }\href {\doibase 10.1103/PhysRevA.38.2423} {\bibfield  {journal} {\bibinfo
  {journal} {Physical Review A}\ }\textbf {\bibinfo {volume} {38}},\ \bibinfo
  {pages} {2423} (\bibinfo {year} {1988})}\BibitemShut {NoStop}%
\bibitem [{\citenamefont {Albert}\ \emph {et~al.}(2011)\citenamefont {Albert},
  \citenamefont {Flindt},\ and\ \citenamefont {Büttiker}}]{albert2011}%
  \BibitemOpen
  \bibfield  {author} {\bibinfo {author} {\bibfnamefont {M.}~\bibnamefont
  {Albert}}, \bibinfo {author} {\bibfnamefont {C.}~\bibnamefont {Flindt}}, \
  and\ \bibinfo {author} {\bibfnamefont {M.}~\bibnamefont {Büttiker}},\ }\href
  {\doibase 10.1103/PhysRevLett.107.086805} {\bibfield  {journal} {\bibinfo
  {journal} {Physical Review Letters}\ }\textbf {\bibinfo {volume} {107}},\
  \bibinfo {pages} {086805} (\bibinfo {year} {2011})},\ \bibinfo {note} {arXiv:
  1102.4452}\BibitemShut {NoStop}%
\bibitem [{\citenamefont {Albert}\ \emph {et~al.}(2012)\citenamefont {Albert},
  \citenamefont {Haack}, \citenamefont {Flindt},\ and\ \citenamefont
  {Büttiker}}]{albert2012}%
  \BibitemOpen
  \bibfield  {author} {\bibinfo {author} {\bibfnamefont {M.}~\bibnamefont
  {Albert}}, \bibinfo {author} {\bibfnamefont {G.}~\bibnamefont {Haack}},
  \bibinfo {author} {\bibfnamefont {C.}~\bibnamefont {Flindt}}, \ and\ \bibinfo
  {author} {\bibfnamefont {M.}~\bibnamefont {Büttiker}},\ }\href {\doibase
  10.1103/PhysRevLett.108.186806} {\bibfield  {journal} {\bibinfo  {journal}
  {Physical Review Letters}\ }\textbf {\bibinfo {volume} {108}},\ \bibinfo
  {pages} {186806} (\bibinfo {year} {2012})},\ \bibinfo {note} {arXiv:
  1202.3152}\BibitemShut {NoStop}%
\bibitem [{\citenamefont {Thomas}\ and\ \citenamefont
  {Flindt}(2013)}]{thomas2013}%
  \BibitemOpen
  \bibfield  {author} {\bibinfo {author} {\bibfnamefont {K.~H.}\ \bibnamefont
  {Thomas}}\ and\ \bibinfo {author} {\bibfnamefont {C.}~\bibnamefont
  {Flindt}},\ }\href {\doibase 10.1103/PhysRevB.87.121405} {\bibfield
  {journal} {\bibinfo  {journal} {Physical Review B}\ }\textbf {\bibinfo
  {volume} {87}},\ \bibinfo {pages} {121405} (\bibinfo {year} {2013})},\
  \bibinfo {note} {arXiv: 1211.4995v1}\BibitemShut {NoStop}%
\bibitem [{\citenamefont {Rajabi}\ \emph {et~al.}(2013)\citenamefont {Rajabi},
  \citenamefont {Pöltl},\ and\ \citenamefont {Governale}}]{rajabi2013}%
  \BibitemOpen
  \bibfield  {author} {\bibinfo {author} {\bibfnamefont {L.}~\bibnamefont
  {Rajabi}}, \bibinfo {author} {\bibfnamefont {C.}~\bibnamefont {Pöltl}}, \
  and\ \bibinfo {author} {\bibfnamefont {M.}~\bibnamefont {Governale}},\ }\href
  {\doibase 10.1103/PhysRevLett.111.067002} {\bibfield  {journal} {\bibinfo
  {journal} {Physical Review Letters}\ }\textbf {\bibinfo {volume} {111}},\
  \bibinfo {pages} {067002} (\bibinfo {year} {2013})},\ \bibinfo {note} {arXiv:
  1304.4301v2}\BibitemShut {NoStop}%
\bibitem [{\citenamefont {Haack}\ \emph {et~al.}(2014)\citenamefont {Haack},
  \citenamefont {Albert},\ and\ \citenamefont {Flindt}}]{haack2014}%
  \BibitemOpen
  \bibfield  {author} {\bibinfo {author} {\bibfnamefont {G.}~\bibnamefont
  {Haack}}, \bibinfo {author} {\bibfnamefont {M.}~\bibnamefont {Albert}}, \
  and\ \bibinfo {author} {\bibfnamefont {C.}~\bibnamefont {Flindt}},\ }\href
  {\doibase 10.1103/PhysRevB.90.205429} {\bibfield  {journal} {\bibinfo
  {journal} {Physical Review B}\ }\textbf {\bibinfo {volume} {90}},\ \bibinfo
  {pages} {205429} (\bibinfo {year} {2014})},\ \bibinfo {note} {arXiv:
  1408.6100}\BibitemShut {NoStop}%
\bibitem [{\citenamefont {Thomas}\ and\ \citenamefont
  {Flindt}(2014)}]{thomas2014}%
  \BibitemOpen
  \bibfield  {author} {\bibinfo {author} {\bibfnamefont {K.~H.}\ \bibnamefont
  {Thomas}}\ and\ \bibinfo {author} {\bibfnamefont {C.}~\bibnamefont
  {Flindt}},\ }\href {\doibase 10.1103/PhysRevB.89.245420} {\bibfield
  {journal} {\bibinfo  {journal} {Physical Review B}\ }\textbf {\bibinfo
  {volume} {89}},\ \bibinfo {pages} {245420} (\bibinfo {year} {2014})},\
  \bibinfo {note} {arXiv: 1402.5033}\BibitemShut {NoStop}%
\bibitem [{\citenamefont {Dasenbrook}\ \emph {et~al.}(2015)\citenamefont
  {Dasenbrook}, \citenamefont {Hofer},\ and\ \citenamefont
  {Flindt}}]{dasenbrook2015}%
  \BibitemOpen
  \bibfield  {author} {\bibinfo {author} {\bibfnamefont {D.}~\bibnamefont
  {Dasenbrook}}, \bibinfo {author} {\bibfnamefont {P.~P.}\ \bibnamefont
  {Hofer}}, \ and\ \bibinfo {author} {\bibfnamefont {C.}~\bibnamefont
  {Flindt}},\ }\href {\doibase 10.1103/PhysRevB.91.195420} {\bibfield
  {journal} {\bibinfo  {journal} {Physical Review B}\ }\textbf {\bibinfo
  {volume} {91}},\ \bibinfo {pages} {195420} (\bibinfo {year} {2015})},\
  \bibinfo {note} {arXiv: 1503.04076}\BibitemShut {NoStop}%
\bibitem [{\citenamefont {Pollock}\ \emph
  {et~al.}(2018{\natexlab{a}})\citenamefont {Pollock}, \citenamefont
  {Rodríguez-Rosario}, \citenamefont {Frauenheim}, \citenamefont
  {Paternostro},\ and\ \citenamefont {Modi}}]{pollock2018a}%
  \BibitemOpen
  \bibfield  {author} {\bibinfo {author} {\bibfnamefont {F.~A.}\ \bibnamefont
  {Pollock}}, \bibinfo {author} {\bibfnamefont {C.}~\bibnamefont
  {Rodríguez-Rosario}}, \bibinfo {author} {\bibfnamefont {T.}~\bibnamefont
  {Frauenheim}}, \bibinfo {author} {\bibfnamefont {M.}~\bibnamefont
  {Paternostro}}, \ and\ \bibinfo {author} {\bibfnamefont {K.}~\bibnamefont
  {Modi}},\ }\href {\doibase 10.1103/PhysRevA.97.012127} {\bibfield  {journal}
  {\bibinfo  {journal} {Physical Review A}\ }\textbf {\bibinfo {volume} {97}},\
  \bibinfo {pages} {012127} (\bibinfo {year} {2018}{\natexlab{a}})},\ \bibinfo
  {note} {arXiv: 1512.00589}\BibitemShut {NoStop}%
\bibitem [{\citenamefont {Pollock}\ \emph
  {et~al.}(2018{\natexlab{b}})\citenamefont {Pollock}, \citenamefont
  {Rodríguez-Rosario}, \citenamefont {Frauenheim}, \citenamefont
  {Paternostro},\ and\ \citenamefont {Modi}}]{pollock2018}%
  \BibitemOpen
  \bibfield  {author} {\bibinfo {author} {\bibfnamefont {F.~A.}\ \bibnamefont
  {Pollock}}, \bibinfo {author} {\bibfnamefont {C.}~\bibnamefont
  {Rodríguez-Rosario}}, \bibinfo {author} {\bibfnamefont {T.}~\bibnamefont
  {Frauenheim}}, \bibinfo {author} {\bibfnamefont {M.}~\bibnamefont
  {Paternostro}}, \ and\ \bibinfo {author} {\bibfnamefont {K.}~\bibnamefont
  {Modi}},\ }\href {\doibase 10.1103/PhysRevLett.120.040405} {\bibfield
  {journal} {\bibinfo  {journal} {Physical Review Letters}\ }\textbf {\bibinfo
  {volume} {120}},\ \bibinfo {pages} {040405} (\bibinfo {year}
  {2018}{\natexlab{b}})},\ \bibinfo {note} {arXiv: 1801.09811}\BibitemShut
  {NoStop}%
\bibitem [{\citenamefont {Milz}\ and\ \citenamefont {Modi}(2021)}]{milz2021}%
  \BibitemOpen
  \bibfield  {author} {\bibinfo {author} {\bibfnamefont {S.}~\bibnamefont
  {Milz}}\ and\ \bibinfo {author} {\bibfnamefont {K.}~\bibnamefont {Modi}},\
  }\href {\doibase 10.1103/PRXQuantum.2.030201} {\bibfield  {journal} {\bibinfo
   {journal} {PRX Quantum}\ }\textbf {\bibinfo {volume} {2}},\ \bibinfo {pages}
  {030201} (\bibinfo {year} {2021})},\ \bibinfo {note} {arXiv:
  2012.01894}\BibitemShut {NoStop}%
\bibitem [{\citenamefont {Liu}\ \emph {et~al.}(2019)\citenamefont {Liu},
  \citenamefont {Zhou},\ and\ \citenamefont {Zhou}}]{liu2019a}%
  \BibitemOpen
  \bibfield  {author} {\bibinfo {author} {\bibfnamefont {F.}~\bibnamefont
  {Liu}}, \bibinfo {author} {\bibfnamefont {X.}~\bibnamefont {Zhou}}, \ and\
  \bibinfo {author} {\bibfnamefont {Z.~W.}\ \bibnamefont {Zhou}},\ }\href
  {\doibase 10.1103/PhysRevA.99.052119} {\bibfield  {journal} {\bibinfo
  {journal} {Physical Review A}\ }\textbf {\bibinfo {volume} {99}},\ \bibinfo
  {pages} {052119} (\bibinfo {year} {2019})},\ \bibinfo {note} {publisher:
  American Physical Society}\BibitemShut {NoStop}%
\bibitem [{\citenamefont {Binder}\ \emph {et~al.}(2018)\citenamefont {Binder},
  \citenamefont {Thompson},\ and\ \citenamefont {Gu}}]{binder2018a}%
  \BibitemOpen
  \bibfield  {author} {\bibinfo {author} {\bibfnamefont {F.~C.}\ \bibnamefont
  {Binder}}, \bibinfo {author} {\bibfnamefont {J.}~\bibnamefont {Thompson}}, \
  and\ \bibinfo {author} {\bibfnamefont {M.}~\bibnamefont {Gu}},\ }\href
  {\doibase 10.1103/PhysRevLett.120.240502} {\bibfield  {journal} {\bibinfo
  {journal} {Physical Review Letters}\ }\textbf {\bibinfo {volume} {120}},\
  \bibinfo {pages} {240502} (\bibinfo {year} {2018})},\ \bibinfo {note}
  {publisher: American Physical Society}\BibitemShut {NoStop}%
\bibitem [{\citenamefont {Liu}\ \emph {et~al.}(2018)\citenamefont {Liu},
  \citenamefont {Elliott}, \citenamefont {Binder}, \citenamefont {Franco},\
  and\ \citenamefont {Gu}}]{liu2018}%
  \BibitemOpen
  \bibfield  {author} {\bibinfo {author} {\bibfnamefont {Q.}~\bibnamefont
  {Liu}}, \bibinfo {author} {\bibfnamefont {T.~J.}\ \bibnamefont {Elliott}},
  \bibinfo {author} {\bibfnamefont {F.~C.}\ \bibnamefont {Binder}}, \bibinfo
  {author} {\bibfnamefont {C.~D.}\ \bibnamefont {Franco}}, \ and\ \bibinfo
  {author} {\bibfnamefont {M.}~\bibnamefont {Gu}},\ }\href@noop {} {\
  (\bibinfo {year} {2018})},\ \bibinfo {note} {arXiv: 1810.09668v1}\BibitemShut
  {NoStop}%
\bibitem [{\citenamefont {Yang}\ \emph {et~al.}(2018)\citenamefont {Yang},
  \citenamefont {Binder}, \citenamefont {Narasimhachar},\ and\ \citenamefont
  {Gu}}]{yang2018}%
  \BibitemOpen
  \bibfield  {author} {\bibinfo {author} {\bibfnamefont {C.}~\bibnamefont
  {Yang}}, \bibinfo {author} {\bibfnamefont {F.~C.}\ \bibnamefont {Binder}},
  \bibinfo {author} {\bibfnamefont {V.}~\bibnamefont {Narasimhachar}}, \ and\
  \bibinfo {author} {\bibfnamefont {M.}~\bibnamefont {Gu}},\ }\href {\doibase
  10.1103/PhysRevLett.121.260602} {\bibfield  {journal} {\bibinfo  {journal}
  {Physical Review Letters}\ }\textbf {\bibinfo {volume} {121}},\ \bibinfo
  {pages} {260602} (\bibinfo {year} {2018})},\ \bibinfo {note} {arXiv:
  1803.08220}\BibitemShut {NoStop}%
\bibitem [{\citenamefont {Landi}\ \emph {et~al.}(2022)\citenamefont {Landi},
  \citenamefont {Poletti},\ and\ \citenamefont {Schaller}}]{landi2022a}%
  \BibitemOpen
  \bibfield  {author} {\bibinfo {author} {\bibfnamefont {G.~T.}\ \bibnamefont
  {Landi}}, \bibinfo {author} {\bibfnamefont {D.}~\bibnamefont {Poletti}}, \
  and\ \bibinfo {author} {\bibfnamefont {G.}~\bibnamefont {Schaller}},\ }\href
  {\doibase 10.1103/RevModPhys.94.045006} {\bibfield  {journal} {\bibinfo
  {journal} {Reviews of Modern Physics}\ }\textbf {\bibinfo {volume} {94}},\
  \bibinfo {pages} {045006} (\bibinfo {year} {2022})},\ \bibinfo {note}
  {publisher: American Physical Society}\BibitemShut {NoStop}%
\bibitem [{\citenamefont {Bertini}\ \emph {et~al.}(2021)\citenamefont
  {Bertini}, \citenamefont {Heidrich-Meisner}, \citenamefont {Karrasch},
  \citenamefont {Prosen}, \citenamefont {Steinigeweg},\ and\ \citenamefont
  {Žnidarič}}]{bertini2021}%
  \BibitemOpen
  \bibfield  {author} {\bibinfo {author} {\bibfnamefont {B.}~\bibnamefont
  {Bertini}}, \bibinfo {author} {\bibfnamefont {F.}~\bibnamefont
  {Heidrich-Meisner}}, \bibinfo {author} {\bibfnamefont {C.}~\bibnamefont
  {Karrasch}}, \bibinfo {author} {\bibfnamefont {T.}~\bibnamefont {Prosen}},
  \bibinfo {author} {\bibfnamefont {R.}~\bibnamefont {Steinigeweg}}, \ and\
  \bibinfo {author} {\bibfnamefont {M.}~\bibnamefont {Žnidarič}},\ }\href
  {\doibase 10.1103/RevModPhys.93.025003} {\bibfield  {journal} {\bibinfo
  {journal} {Reviews of Modern Physics}\ }\textbf {\bibinfo {volume} {93}},\
  \bibinfo {pages} {025003} (\bibinfo {year} {2021})},\ \bibinfo {note} {arXiv:
  2003.03334}\BibitemShut {NoStop}%
\bibitem [{\citenamefont {Prosen}\ and\ \citenamefont
  {Pižorn}(2008)}]{prosen2008}%
  \BibitemOpen
  \bibfield  {author} {\bibinfo {author} {\bibfnamefont {T.}~\bibnamefont
  {Prosen}}\ and\ \bibinfo {author} {\bibfnamefont {I.}~\bibnamefont
  {Pižorn}},\ }\href {\doibase 10.1103/PhysRevLett.101.105701} {\bibfield
  {journal} {\bibinfo  {journal} {Physical Review Letters}\ }\textbf {\bibinfo
  {volume} {101}},\ \bibinfo {pages} {105701} (\bibinfo {year} {2008})},\
  \bibinfo {note} {arXiv: 0805.2878}\BibitemShut {NoStop}%
\bibitem [{\citenamefont {Benenti}\ \emph {et~al.}(2009)\citenamefont
  {Benenti}, \citenamefont {Casati}, \citenamefont {Prosen}, \citenamefont
  {Rossini},\ and\ \citenamefont {Žnidarič}}]{benenti2009}%
  \BibitemOpen
  \bibfield  {author} {\bibinfo {author} {\bibfnamefont {G.}~\bibnamefont
  {Benenti}}, \bibinfo {author} {\bibfnamefont {G.}~\bibnamefont {Casati}},
  \bibinfo {author} {\bibfnamefont {T.}~\bibnamefont {Prosen}}, \bibinfo
  {author} {\bibfnamefont {D.}~\bibnamefont {Rossini}}, \ and\ \bibinfo
  {author} {\bibfnamefont {M.}~\bibnamefont {Žnidarič}},\ }\href {\doibase
  10.1103/PhysRevB.80.035110} {\bibfield  {journal} {\bibinfo  {journal}
  {Physical Review B}\ }\textbf {\bibinfo {volume} {80}},\ \bibinfo {pages}
  {035110} (\bibinfo {year} {2009})}\BibitemShut {NoStop}%
\bibitem [{\citenamefont {Karevski}\ and\ \citenamefont
  {Platini}(2009)}]{karevski2009}%
  \BibitemOpen
  \bibfield  {author} {\bibinfo {author} {\bibfnamefont {D.}~\bibnamefont
  {Karevski}}\ and\ \bibinfo {author} {\bibfnamefont {T.}~\bibnamefont
  {Platini}},\ }\href {\doibase 10.1103/PhysRevLett.102.207207} {\bibfield
  {journal} {\bibinfo  {journal} {Physical Review Letters}\ }\textbf {\bibinfo
  {volume} {102}},\ \bibinfo {pages} {207207} (\bibinfo {year} {2009})},\
  \bibinfo {note} {arXiv: 0904.3527}\BibitemShut {NoStop}%
\bibitem [{\citenamefont {Prosen}\ and\ \citenamefont
  {Zunkovic}(2010)}]{prosen2010}%
  \BibitemOpen
  \bibfield  {author} {\bibinfo {author} {\bibfnamefont {T.}~\bibnamefont
  {Prosen}}\ and\ \bibinfo {author} {\bibfnamefont {B.}~\bibnamefont
  {Zunkovic}},\ }\href {\doibase 10.1088/1367-2630/12/2/025016} {\bibfield
  {journal} {\bibinfo  {journal} {New Journal of Physics}\ }\textbf {\bibinfo
  {volume} {12}},\ \bibinfo {pages} {025016} (\bibinfo {year} {2010})},\
  \bibinfo {note} {arXiv: 0910.0195}\BibitemShut {NoStop}%
\bibitem [{\citenamefont {Dzhioev}\ and\ \citenamefont
  {Kosov}(2011)}]{dzhioev2011}%
  \BibitemOpen
  \bibfield  {author} {\bibinfo {author} {\bibfnamefont {A.~A.}\ \bibnamefont
  {Dzhioev}}\ and\ \bibinfo {author} {\bibfnamefont {D.~S.}\ \bibnamefont
  {Kosov}},\ }\href {\doibase 10.1063/1.3548065} {\bibfield  {journal}
  {\bibinfo  {journal} {Journal of Chemical Physics}\ }\textbf {\bibinfo
  {volume} {134}},\ \bibinfo {pages} {1} (\bibinfo {year} {2011})},\ \bibinfo
  {note} {arXiv: 1007.4643v2}\BibitemShut {NoStop}%
\bibitem [{\citenamefont {Popkov}\ \emph {et~al.}(2012)\citenamefont {Popkov},
  \citenamefont {Salerno},\ and\ \citenamefont {Schütz}}]{popkov2012}%
  \BibitemOpen
  \bibfield  {author} {\bibinfo {author} {\bibfnamefont {V.}~\bibnamefont
  {Popkov}}, \bibinfo {author} {\bibfnamefont {M.}~\bibnamefont {Salerno}}, \
  and\ \bibinfo {author} {\bibfnamefont {G.~M.}\ \bibnamefont {Schütz}},\
  }\href {\doibase 10.1103/PhysRevE.85.031137} {\bibfield  {journal} {\bibinfo
  {journal} {Physical Review E}\ }\textbf {\bibinfo {volume} {85}},\ \bibinfo
  {pages} {031137} (\bibinfo {year} {2012})}\BibitemShut {NoStop}%
\bibitem [{\citenamefont {Mendoza‐Arenas}\ \emph
  {et~al.}(2013{\natexlab{a}})\citenamefont {Mendoza‐Arenas}, \citenamefont
  {Al-Assam}, \citenamefont {Clark},\ and\ \citenamefont
  {Jaksch}}]{mendoza-arenas2013a}%
  \BibitemOpen
  \bibfield  {author} {\bibinfo {author} {\bibfnamefont {J.~J.}\ \bibnamefont
  {Mendoza‐Arenas}}, \bibinfo {author} {\bibfnamefont {S.}~\bibnamefont
  {Al-Assam}}, \bibinfo {author} {\bibfnamefont {S.~R.}\ \bibnamefont {Clark}},
  \ and\ \bibinfo {author} {\bibfnamefont {D.}~\bibnamefont {Jaksch}},\ }\href
  {\doibase 10.1088/1742-5468/2013/07/P07007} {\bibfield  {journal} {\bibinfo
  {journal} {Journal of Statistical Mechanics: Theory and Experiment}\ }\textbf
  {\bibinfo {volume} {2013}},\ \bibinfo {pages} {P07007} (\bibinfo {year}
  {2013}{\natexlab{a}})}\BibitemShut {NoStop}%
\bibitem [{\citenamefont {Mendoza‐Arenas}\ \emph
  {et~al.}(2013{\natexlab{b}})\citenamefont {Mendoza‐Arenas}, \citenamefont
  {Grujic}, \citenamefont {Jaksch},\ and\ \citenamefont
  {Clark}}]{mendoza-arenas2013}%
  \BibitemOpen
  \bibfield  {author} {\bibinfo {author} {\bibfnamefont {J.~J.}\ \bibnamefont
  {Mendoza‐Arenas}}, \bibinfo {author} {\bibfnamefont {T.}~\bibnamefont
  {Grujic}}, \bibinfo {author} {\bibfnamefont {D.}~\bibnamefont {Jaksch}}, \
  and\ \bibinfo {author} {\bibfnamefont {S.~R.}\ \bibnamefont {Clark}},\ }\href
  {\doibase 10.1103/PhysRevB.87.235130} {\bibfield  {journal} {\bibinfo
  {journal} {Physical Review B}\ }\textbf {\bibinfo {volume} {87}},\ \bibinfo
  {pages} {235130} (\bibinfo {year} {2013}{\natexlab{b}})}\BibitemShut
  {NoStop}%
\bibitem [{\citenamefont {Novotný}\ \emph {et~al.}(2003)\citenamefont
  {Novotný}, \citenamefont {Donarini},\ and\ \citenamefont
  {Jauho}}]{novotny2003}%
  \BibitemOpen
  \bibfield  {author} {\bibinfo {author} {\bibfnamefont {T.}~\bibnamefont
  {Novotný}}, \bibinfo {author} {\bibfnamefont {A.}~\bibnamefont {Donarini}},
  \ and\ \bibinfo {author} {\bibfnamefont {A.-P.}\ \bibnamefont {Jauho}},\
  }\href {\doibase 10.1103/PhysRevLett.90.256801} {\bibfield  {journal}
  {\bibinfo  {journal} {Physical Review Letters}\ }\textbf {\bibinfo {volume}
  {90}},\ \bibinfo {pages} {256801} (\bibinfo {year} {2003})},\ \bibinfo {note}
  {publisher: American Physical Society}\BibitemShut {NoStop}%
\bibitem [{\citenamefont {Flindt}\ \emph {et~al.}(2004)\citenamefont {Flindt},
  \citenamefont {Novotný},\ and\ \citenamefont {Jauho}}]{flindt2004a}%
  \BibitemOpen
  \bibfield  {author} {\bibinfo {author} {\bibfnamefont {C.}~\bibnamefont
  {Flindt}}, \bibinfo {author} {\bibfnamefont {T.}~\bibnamefont {Novotný}}, \
  and\ \bibinfo {author} {\bibfnamefont {A.-P.}\ \bibnamefont {Jauho}},\ }\href
  {\doibase 10.1103/PhysRevB.70.205334} {\bibfield  {journal} {\bibinfo
  {journal} {Physical Review B}\ }\textbf {\bibinfo {volume} {70}},\ \bibinfo
  {pages} {205334} (\bibinfo {year} {2004})}\BibitemShut {NoStop}%
\bibitem [{\citenamefont {Minganti}\ \emph {et~al.}(2018)\citenamefont
  {Minganti}, \citenamefont {Biella}, \citenamefont {Bartolo},\ and\
  \citenamefont {Ciuti}}]{minganti2018}%
  \BibitemOpen
  \bibfield  {author} {\bibinfo {author} {\bibfnamefont {F.}~\bibnamefont
  {Minganti}}, \bibinfo {author} {\bibfnamefont {A.}~\bibnamefont {Biella}},
  \bibinfo {author} {\bibfnamefont {N.}~\bibnamefont {Bartolo}}, \ and\
  \bibinfo {author} {\bibfnamefont {C.}~\bibnamefont {Ciuti}},\ }\href
  {http://arxiv.org/abs/1804.11293} {\bibfield  {journal} {\bibinfo  {journal}
  {Physical Review A}\ }\textbf {\bibinfo {volume} {98}},\ \bibinfo {pages}
  {042118} (\bibinfo {year} {2018})},\ \bibinfo {note} {arXiv:
  1804.11293}\BibitemShut {NoStop}%
\bibitem [{\citenamefont {Guo}\ and\ \citenamefont {Poletti}(2017)}]{guo2017}%
  \BibitemOpen
  \bibfield  {author} {\bibinfo {author} {\bibfnamefont {C.}~\bibnamefont
  {Guo}}\ and\ \bibinfo {author} {\bibfnamefont {D.}~\bibnamefont {Poletti}},\
  }\href {\doibase 10.1103/PhysRevA.95.052107} {\bibfield  {journal} {\bibinfo
  {journal} {Physical Review A}\ }\textbf {\bibinfo {volume} {95}},\ \bibinfo
  {pages} {052107} (\bibinfo {year} {2017})}\BibitemShut {NoStop}%
\bibitem [{\citenamefont {Landi}\ \emph {et~al.}(2014)\citenamefont {Landi},
  \citenamefont {Novais}, \citenamefont {de~Oliveira},\ and\ \citenamefont
  {Karevski}}]{landi2014}%
  \BibitemOpen
  \bibfield  {author} {\bibinfo {author} {\bibfnamefont {G.~T.}\ \bibnamefont
  {Landi}}, \bibinfo {author} {\bibfnamefont {E.}~\bibnamefont {Novais}},
  \bibinfo {author} {\bibfnamefont {M.~J.}\ \bibnamefont {de~Oliveira}}, \ and\
  \bibinfo {author} {\bibfnamefont {D.}~\bibnamefont {Karevski}},\ }\href
  {\doibase 10.1103/PhysRevE.90.042142} {\bibfield  {journal} {\bibinfo
  {journal} {Physical Review E}\ }\textbf {\bibinfo {volume} {90}},\ \bibinfo
  {pages} {042142} (\bibinfo {year} {2014})}\BibitemShut {NoStop}%
\bibitem [{\citenamefont {Landi}\ and\ \citenamefont
  {Karevski}(2015)}]{landi2015}%
  \BibitemOpen
  \bibfield  {author} {\bibinfo {author} {\bibfnamefont {G.~T.}\ \bibnamefont
  {Landi}}\ and\ \bibinfo {author} {\bibfnamefont {D.}~\bibnamefont
  {Karevski}},\ }\href {\doibase 10.1103/PhysRevB.91.174422} {\bibfield
  {journal} {\bibinfo  {journal} {Physical Review B}\ }\textbf {\bibinfo
  {volume} {91}},\ \bibinfo {pages} {174422} (\bibinfo {year}
  {2015})}\BibitemShut {NoStop}%
\bibitem [{\citenamefont {Thingna}\ and\ \citenamefont
  {Wang}(2013)}]{thingna2013}%
  \BibitemOpen
  \bibfield  {author} {\bibinfo {author} {\bibfnamefont {J.}~\bibnamefont
  {Thingna}}\ and\ \bibinfo {author} {\bibfnamefont {J.-S.}\ \bibnamefont
  {Wang}},\ }\href {\doibase 10.1209/0295-5075/104/37006} {\bibfield  {journal}
  {\bibinfo  {journal} {EPL (Europhysics Letters)}\ }\textbf {\bibinfo {volume}
  {104}},\ \bibinfo {pages} {37006} (\bibinfo {year} {2013})}\BibitemShut
  {NoStop}%
\bibitem [{\citenamefont {Balachandran}\ \emph {et~al.}(2018)\citenamefont
  {Balachandran}, \citenamefont {Benenti}, \citenamefont {Pereira},
  \citenamefont {Casati},\ and\ \citenamefont {Poletti}}]{balachandran2018}%
  \BibitemOpen
  \bibfield  {author} {\bibinfo {author} {\bibfnamefont {V.}~\bibnamefont
  {Balachandran}}, \bibinfo {author} {\bibfnamefont {G.}~\bibnamefont
  {Benenti}}, \bibinfo {author} {\bibfnamefont {E.}~\bibnamefont {Pereira}},
  \bibinfo {author} {\bibfnamefont {G.}~\bibnamefont {Casati}}, \ and\ \bibinfo
  {author} {\bibfnamefont {D.}~\bibnamefont {Poletti}},\ }\href {\doibase
  10.1103/PhysRevLett.120.200603} {\bibfield  {journal} {\bibinfo  {journal}
  {Physical Review Letters}\ }\textbf {\bibinfo {volume} {120}},\ \bibinfo
  {pages} {200603} (\bibinfo {year} {2018})},\ \bibinfo {note} {arXiv:
  1707.08823 Publisher: American Physical Society}\BibitemShut {NoStop}%
\bibitem [{\citenamefont {Balachandran}\ \emph {et~al.}(2019)\citenamefont
  {Balachandran}, \citenamefont {Clark}, \citenamefont {Goold},\ and\
  \citenamefont {Poletti}}]{balachandran2019}%
  \BibitemOpen
  \bibfield  {author} {\bibinfo {author} {\bibfnamefont {V.}~\bibnamefont
  {Balachandran}}, \bibinfo {author} {\bibfnamefont {S.~R.}\ \bibnamefont
  {Clark}}, \bibinfo {author} {\bibfnamefont {J.}~\bibnamefont {Goold}}, \ and\
  \bibinfo {author} {\bibfnamefont {D.}~\bibnamefont {Poletti}},\ }\href
  {\doibase 10.1103/PhysRevLett.123.020603} {\bibfield  {journal} {\bibinfo
  {journal} {Physical Review Letters}\ }\textbf {\bibinfo {volume} {123}},\
  \bibinfo {pages} {020603} (\bibinfo {year} {2019})},\ \bibinfo {note}
  {publisher: American Physical Society}\BibitemShut {NoStop}%
\bibitem [{\citenamefont {Mendoza-Arenas}\ and\ \citenamefont
  {Clark}(2022)}]{mendoza-arenas2022}%
  \BibitemOpen
  \bibfield  {author} {\bibinfo {author} {\bibfnamefont {J.~J.}\ \bibnamefont
  {Mendoza-Arenas}}\ and\ \bibinfo {author} {\bibfnamefont {S.~R.}\
  \bibnamefont {Clark}},\ }\href {http://arxiv.org/abs/2209.11718} {\enquote
  {\bibinfo {title} {Giant rectification in strongly-interacting
  boundary-driven tilted systems},}\ } (\bibinfo {year} {2022}),\ \bibinfo
  {note} {arXiv:2209.11718 [cond-mat, physics:quant-ph]}\BibitemShut {NoStop}%
\bibitem [{\citenamefont {Rebentrost}\ \emph {et~al.}(2009)\citenamefont
  {Rebentrost}, \citenamefont {Mohseni}, \citenamefont {Kassal}, \citenamefont
  {Lloyd},\ and\ \citenamefont {Aspuru-Guzik}}]{rebentrost2009}%
  \BibitemOpen
  \bibfield  {author} {\bibinfo {author} {\bibfnamefont {P.}~\bibnamefont
  {Rebentrost}}, \bibinfo {author} {\bibfnamefont {M.}~\bibnamefont {Mohseni}},
  \bibinfo {author} {\bibfnamefont {I.}~\bibnamefont {Kassal}}, \bibinfo
  {author} {\bibfnamefont {S.}~\bibnamefont {Lloyd}}, \ and\ \bibinfo {author}
  {\bibfnamefont {A.}~\bibnamefont {Aspuru-Guzik}},\ }\href {\doibase
  10.1088/1367-2630/11/3/033003} {\bibfield  {journal} {\bibinfo  {journal}
  {New Journal of Physics}\ }\textbf {\bibinfo {volume} {11}},\ \bibinfo
  {pages} {033003} (\bibinfo {year} {2009})},\ \bibinfo {note} {arXiv:
  0807.0929}\BibitemShut {NoStop}%
\bibitem [{\citenamefont {Chiaracane}\ \emph {et~al.}(2021)\citenamefont
  {Chiaracane}, \citenamefont {Pietracaprina}, \citenamefont {Purkayastha},\
  and\ \citenamefont {Goold}}]{chiaracane2021a}%
  \BibitemOpen
  \bibfield  {author} {\bibinfo {author} {\bibfnamefont {C.}~\bibnamefont
  {Chiaracane}}, \bibinfo {author} {\bibfnamefont {F.}~\bibnamefont
  {Pietracaprina}}, \bibinfo {author} {\bibfnamefont {A.}~\bibnamefont
  {Purkayastha}}, \ and\ \bibinfo {author} {\bibfnamefont {J.}~\bibnamefont
  {Goold}},\ }\href {http://arxiv.org/abs/2101.01111} {\ ,\ \bibinfo {pages}
  {1} (\bibinfo {year} {2021})},\ \bibinfo {note} {arXiv:
  2101.01111}\BibitemShut {NoStop}%
\bibitem [{\citenamefont {Lacerda}\ \emph {et~al.}(2021)\citenamefont
  {Lacerda}, \citenamefont {Goold},\ and\ \citenamefont
  {Landi}}]{lacerda2021a}%
  \BibitemOpen
  \bibfield  {author} {\bibinfo {author} {\bibfnamefont {A.~M.}\ \bibnamefont
  {Lacerda}}, \bibinfo {author} {\bibfnamefont {J.}~\bibnamefont {Goold}}, \
  and\ \bibinfo {author} {\bibfnamefont {G.~T.}\ \bibnamefont {Landi}},\ }\href
  {\doibase 10.1103/PhysRevB.104.174203} {\bibfield  {journal} {\bibinfo
  {journal} {Physical Review B}\ }\textbf {\bibinfo {volume} {104}},\ \bibinfo
  {pages} {174203} (\bibinfo {year} {2021})},\ \bibinfo {note} {publisher:
  American Physical Society}\BibitemShut {NoStop}%
\bibitem [{\citenamefont {Varma}\ and\ \citenamefont
  {Žnidarič}(2019)}]{varma2019a}%
  \BibitemOpen
  \bibfield  {author} {\bibinfo {author} {\bibfnamefont {V.~K.}\ \bibnamefont
  {Varma}}\ and\ \bibinfo {author} {\bibfnamefont {M.}~\bibnamefont
  {Žnidarič}},\ }\href {\doibase 10.1103/PhysRevB.100.085105} {\bibfield
  {journal} {\bibinfo  {journal} {Physical Review B}\ }\textbf {\bibinfo
  {volume} {100}},\ \bibinfo {pages} {085105} (\bibinfo {year} {2019})},\
  \bibinfo {note} {publisher: American Physical Society}\BibitemShut {NoStop}%
\bibitem [{\citenamefont {Schaller}(2014)}]{schaller2014}%
  \BibitemOpen
  \bibfield  {author} {\bibinfo {author} {\bibfnamefont {G.}~\bibnamefont
  {Schaller}},\ }\href {\doibase 10.1007/978-3-319-03877-3} {\emph {\bibinfo
  {title} {Open {Quantum} {Systems} {Far} from {Equilibrium}}}}\ (\bibinfo
  {year} {2014})\BibitemShut {NoStop}%
\bibitem [{\citenamefont {Esposito}\ \emph {et~al.}(2009)\citenamefont
  {Esposito}, \citenamefont {Harbola},\ and\ \citenamefont
  {Mukamel}}]{esposito2009}%
  \BibitemOpen
  \bibfield  {author} {\bibinfo {author} {\bibfnamefont {M.}~\bibnamefont
  {Esposito}}, \bibinfo {author} {\bibfnamefont {U.}~\bibnamefont {Harbola}}, \
  and\ \bibinfo {author} {\bibfnamefont {S.}~\bibnamefont {Mukamel}},\ }\href
  {\doibase 10.1103/RevModPhys.81.1665} {\bibfield  {journal} {\bibinfo
  {journal} {Reviews of Modern Physics}\ }\textbf {\bibinfo {volume} {81}},\
  \bibinfo {pages} {1665} (\bibinfo {year} {2009})}\BibitemShut {NoStop}%
\bibitem [{\citenamefont {Levitov}\ and\ \citenamefont
  {Lesovik}(1993)}]{levitov1993}%
  \BibitemOpen
  \bibfield  {author} {\bibinfo {author} {\bibfnamefont {L.}~\bibnamefont
  {Levitov}}\ and\ \bibinfo {author} {\bibfnamefont {G.}~\bibnamefont
  {Lesovik}},\ }\href@noop {} {\bibfield  {journal} {\bibinfo  {journal} {JETP
  letters}\ }\textbf {\bibinfo {volume} {58}},\ \bibinfo {pages} {230}
  (\bibinfo {year} {1993})}\BibitemShut {NoStop}%
\bibitem [{\citenamefont {Levitov}\ \emph {et~al.}(1996)\citenamefont
  {Levitov}, \citenamefont {Lee},\ and\ \citenamefont {Lesovik}}]{levitov1996}%
  \BibitemOpen
  \bibfield  {author} {\bibinfo {author} {\bibfnamefont {L.~S.}\ \bibnamefont
  {Levitov}}, \bibinfo {author} {\bibfnamefont {H.}~\bibnamefont {Lee}}, \ and\
  \bibinfo {author} {\bibfnamefont {G.~B.}\ \bibnamefont {Lesovik}},\ }\href
  {\doibase 10.1063/1.531672} {\bibfield  {journal} {\bibinfo  {journal}
  {Journal of Mathematical Physics}\ }\textbf {\bibinfo {volume} {37}},\
  \bibinfo {pages} {4845} (\bibinfo {year} {1996})}\BibitemShut {NoStop}%
\bibitem [{\citenamefont {Nazarov}\ and\ \citenamefont
  {Kindermann}(2003)}]{nazarov2003}%
  \BibitemOpen
  \bibfield  {author} {\bibinfo {author} {\bibfnamefont {Y.~V.}\ \bibnamefont
  {Nazarov}}\ and\ \bibinfo {author} {\bibfnamefont {M.}~\bibnamefont
  {Kindermann}},\ }\href {\doibase 10.1140/epjb/e2003-00293-1} {\bibfield
  {journal} {\bibinfo  {journal} {The European Physical Journal B - Condensed
  Matter and Complex Systems}\ }\textbf {\bibinfo {volume} {35}},\ \bibinfo
  {pages} {413} (\bibinfo {year} {2003})}\BibitemShut {NoStop}%
\bibitem [{\citenamefont {Flindt}\ \emph {et~al.}(2010)\citenamefont {Flindt},
  \citenamefont {Novotný}, \citenamefont {Braggio},\ and\ \citenamefont
  {Jauho}}]{flindt2010}%
  \BibitemOpen
  \bibfield  {author} {\bibinfo {author} {\bibfnamefont {C.}~\bibnamefont
  {Flindt}}, \bibinfo {author} {\bibfnamefont {T.}~\bibnamefont {Novotný}},
  \bibinfo {author} {\bibfnamefont {A.}~\bibnamefont {Braggio}}, \ and\
  \bibinfo {author} {\bibfnamefont {A.-P.}\ \bibnamefont {Jauho}},\ }\href
  {\doibase 10.1103/PhysRevB.82.155407} {\bibfield  {journal} {\bibinfo
  {journal} {Physical Review B}\ }\textbf {\bibinfo {volume} {82}},\ \bibinfo
  {pages} {155407} (\bibinfo {year} {2010})},\ \bibinfo {note} {arXiv:
  1002.4506}\BibitemShut {NoStop}%
\bibitem [{\citenamefont {Budini}\ \emph {et~al.}(2014)\citenamefont {Budini},
  \citenamefont {Turner},\ and\ \citenamefont {Garrahan}}]{budini2014}%
  \BibitemOpen
  \bibfield  {author} {\bibinfo {author} {\bibfnamefont {A.~A.}\ \bibnamefont
  {Budini}}, \bibinfo {author} {\bibfnamefont {R.~M.}\ \bibnamefont {Turner}},
  \ and\ \bibinfo {author} {\bibfnamefont {J.~P.}\ \bibnamefont {Garrahan}},\
  }\href {\doibase 10.1088/1742-5468/2014/03/P03012} {\bibfield  {journal}
  {\bibinfo  {journal} {Journal of Statistical Mechanics: Theory and
  Experiment}\ }\textbf {\bibinfo {volume} {2014}},\ \bibinfo {pages} {P03012}
  (\bibinfo {year} {2014})},\ \bibinfo {note} {publisher: IOP Publishing and
  SISSA}\BibitemShut {NoStop}%
\bibitem [{\citenamefont {Kiukas}\ \emph {et~al.}(2015)\citenamefont {Kiukas},
  \citenamefont {Guţă}, \citenamefont {Lesanovsky},\ and\ \citenamefont
  {Garrahan}}]{kiukas2015}%
  \BibitemOpen
  \bibfield  {author} {\bibinfo {author} {\bibfnamefont {J.}~\bibnamefont
  {Kiukas}}, \bibinfo {author} {\bibfnamefont {M.}~\bibnamefont {Guţă}},
  \bibinfo {author} {\bibfnamefont {I.}~\bibnamefont {Lesanovsky}}, \ and\
  \bibinfo {author} {\bibfnamefont {J.~P.}\ \bibnamefont {Garrahan}},\ }\href
  {\doibase 10.1103/PhysRevE.92.012132} {\bibfield  {journal} {\bibinfo
  {journal} {Physical Review E}\ }\textbf {\bibinfo {volume} {92}},\ \bibinfo
  {pages} {012132} (\bibinfo {year} {2015})}\BibitemShut {NoStop}%
\bibitem [{\citenamefont {Radaelli}\ \emph {et~al.}(2023)\citenamefont
  {Radaelli}, \citenamefont {Landi},\ and\ \citenamefont
  {Binder}}]{radaelli2023}%
  \BibitemOpen
  \bibfield  {author} {\bibinfo {author} {\bibfnamefont {M.}~\bibnamefont
  {Radaelli}}, \bibinfo {author} {\bibfnamefont {G.~T.}\ \bibnamefont {Landi}},
  \ and\ \bibinfo {author} {\bibfnamefont {F.~C.}\ \bibnamefont {Binder}},\
  }\href {http://arxiv.org/abs/2303.15405} {\enquote {\bibinfo {title} {A
  {Gillespie} algorithm for efficient simulation of quantum jump
  trajectories},}\ } (\bibinfo {year} {2023}),\ \bibinfo {note}
  {arXiv:2303.15405 [cond-mat, physics:quant-ph]}\BibitemShut {NoStop}%
\bibitem [{\citenamefont {Turkeshi}\ \emph {et~al.}(2021)\citenamefont
  {Turkeshi}, \citenamefont {Biella}, \citenamefont {Fazio}, \citenamefont
  {Dalmonte},\ and\ \citenamefont {Schiró}}]{turkeshi2021a}%
  \BibitemOpen
  \bibfield  {author} {\bibinfo {author} {\bibfnamefont {X.}~\bibnamefont
  {Turkeshi}}, \bibinfo {author} {\bibfnamefont {A.}~\bibnamefont {Biella}},
  \bibinfo {author} {\bibfnamefont {R.}~\bibnamefont {Fazio}}, \bibinfo
  {author} {\bibfnamefont {M.}~\bibnamefont {Dalmonte}}, \ and\ \bibinfo
  {author} {\bibfnamefont {M.}~\bibnamefont {Schiró}},\ }\href {\doibase
  10.1103/PhysRevB.103.224210} {\bibfield  {journal} {\bibinfo  {journal}
  {Physical Review B}\ }\textbf {\bibinfo {volume} {103}},\ \bibinfo {pages}
  {224210} (\bibinfo {year} {2021})}\BibitemShut {NoStop}%
\bibitem [{\citenamefont {Skinner}\ \emph {et~al.}(2019)\citenamefont
  {Skinner}, \citenamefont {Ruhman},\ and\ \citenamefont
  {Nahum}}]{skinner2019}%
  \BibitemOpen
  \bibfield  {author} {\bibinfo {author} {\bibfnamefont {B.}~\bibnamefont
  {Skinner}}, \bibinfo {author} {\bibfnamefont {J.}~\bibnamefont {Ruhman}}, \
  and\ \bibinfo {author} {\bibfnamefont {A.}~\bibnamefont {Nahum}},\ }\href
  {\doibase 10.1103/PhysRevX.9.031009} {\bibfield  {journal} {\bibinfo
  {journal} {Physical Review X}\ }\textbf {\bibinfo {volume} {9}},\ \bibinfo
  {pages} {031009} (\bibinfo {year} {2019})},\ \bibinfo {note} {publisher:
  American Physical Society}\BibitemShut {NoStop}%
\bibitem [{\citenamefont {Li}\ \emph {et~al.}(2018)\citenamefont {Li},
  \citenamefont {Chen},\ and\ \citenamefont {Fisher}}]{li2018a}%
  \BibitemOpen
  \bibfield  {author} {\bibinfo {author} {\bibfnamefont {Y.}~\bibnamefont
  {Li}}, \bibinfo {author} {\bibfnamefont {X.}~\bibnamefont {Chen}}, \ and\
  \bibinfo {author} {\bibfnamefont {M.~P.~A.}\ \bibnamefont {Fisher}},\ }\href
  {\doibase 10.1103/PhysRevB.98.205136} {\bibfield  {journal} {\bibinfo
  {journal} {Physical Review B}\ }\textbf {\bibinfo {volume} {98}},\ \bibinfo
  {pages} {205136} (\bibinfo {year} {2018})},\ \bibinfo {note} {publisher:
  American Physical Society}\BibitemShut {NoStop}%
\bibitem [{\citenamefont {Coppola}\ \emph {et~al.}(2022)\citenamefont
  {Coppola}, \citenamefont {Tirrito}, \citenamefont {Karevski},\ and\
  \citenamefont {Collura}}]{coppola2022}%
  \BibitemOpen
  \bibfield  {author} {\bibinfo {author} {\bibfnamefont {M.}~\bibnamefont
  {Coppola}}, \bibinfo {author} {\bibfnamefont {E.}~\bibnamefont {Tirrito}},
  \bibinfo {author} {\bibfnamefont {D.}~\bibnamefont {Karevski}}, \ and\
  \bibinfo {author} {\bibfnamefont {M.}~\bibnamefont {Collura}},\ }\href
  {\doibase 10.1103/PhysRevB.105.094303} {\bibfield  {journal} {\bibinfo
  {journal} {Physical Review B}\ }\textbf {\bibinfo {volume} {105}},\ \bibinfo
  {pages} {094303} (\bibinfo {year} {2022})},\ \bibinfo {note} {publisher:
  American Physical Society}\BibitemShut {NoStop}%
\bibitem [{\citenamefont {Alberton}\ \emph {et~al.}(2021)\citenamefont
  {Alberton}, \citenamefont {Buchhold},\ and\ \citenamefont
  {Diehl}}]{alberton2021}%
  \BibitemOpen
  \bibfield  {author} {\bibinfo {author} {\bibfnamefont {O.}~\bibnamefont
  {Alberton}}, \bibinfo {author} {\bibfnamefont {M.}~\bibnamefont {Buchhold}},
  \ and\ \bibinfo {author} {\bibfnamefont {S.}~\bibnamefont {Diehl}},\ }\href
  {\doibase 10.1103/PhysRevLett.126.170602} {\bibfield  {journal} {\bibinfo
  {journal} {Physical Review Letters}\ }\textbf {\bibinfo {volume} {126}},\
  \bibinfo {pages} {170602} (\bibinfo {year} {2021})}\BibitemShut {NoStop}%
\bibitem [{\citenamefont {Carollo}\ and\ \citenamefont
  {Alba}(2022)}]{carollo2022}%
  \BibitemOpen
  \bibfield  {author} {\bibinfo {author} {\bibfnamefont {F.}~\bibnamefont
  {Carollo}}\ and\ \bibinfo {author} {\bibfnamefont {V.}~\bibnamefont {Alba}},\
  }\href {\doibase 10.1103/PhysRevB.106.L220304} {\bibfield  {journal}
  {\bibinfo  {journal} {Physical Review B}\ }\textbf {\bibinfo {volume}
  {106}},\ \bibinfo {pages} {L220304} (\bibinfo {year} {2022})}\BibitemShut
  {NoStop}%
\bibitem [{\citenamefont {Cao}\ \emph {et~al.}(2019)\citenamefont {Cao},
  \citenamefont {Tilloy},\ and\ \citenamefont {De~Luca}}]{cao2019}%
  \BibitemOpen
  \bibfield  {author} {\bibinfo {author} {\bibfnamefont {X.}~\bibnamefont
  {Cao}}, \bibinfo {author} {\bibfnamefont {A.}~\bibnamefont {Tilloy}}, \ and\
  \bibinfo {author} {\bibfnamefont {A.}~\bibnamefont {De~Luca}},\ }\href
  {\doibase 10.21468/SciPostPhys.7.2.024} {\bibfield  {journal} {\bibinfo
  {journal} {SciPost Physics}\ }\textbf {\bibinfo {volume} {7}},\ \bibinfo
  {pages} {024} (\bibinfo {year} {2019})}\BibitemShut {NoStop}%
\bibitem [{\citenamefont {Crutchfield}\ and\ \citenamefont
  {Young}(1989)}]{crutchfield1989}%
  \BibitemOpen
  \bibfield  {author} {\bibinfo {author} {\bibfnamefont {J.~P.}\ \bibnamefont
  {Crutchfield}}\ and\ \bibinfo {author} {\bibfnamefont {K.}~\bibnamefont
  {Young}},\ }\href {\doibase 10.1103/PhysRevLett.63.105} {\bibfield  {journal}
  {\bibinfo  {journal} {Physical Review Letters}\ }\textbf {\bibinfo {volume}
  {63}},\ \bibinfo {pages} {105} (\bibinfo {year} {1989})}\BibitemShut
  {NoStop}%
\bibitem [{\citenamefont {Shalizi}\ and\ \citenamefont
  {Crutchfield}(2001)}]{shalizi2001}%
  \BibitemOpen
  \bibfield  {author} {\bibinfo {author} {\bibfnamefont {C.~R.}\ \bibnamefont
  {Shalizi}}\ and\ \bibinfo {author} {\bibfnamefont {J.~P.}\ \bibnamefont
  {Crutchfield}},\ }\href {\doibase 10.1023/A:1010388907793} {\bibfield
  {journal} {\bibinfo  {journal} {Journal of Statistical Physics}\ }\textbf
  {\bibinfo {volume} {104}},\ \bibinfo {pages} {817} (\bibinfo {year}
  {2001})}\BibitemShut {NoStop}%
\bibitem [{\citenamefont {Shalizi}\ and\ \citenamefont
  {Shalizi}(2004)}]{shalizi2004}%
  \BibitemOpen
  \bibfield  {author} {\bibinfo {author} {\bibfnamefont {C.~R.}\ \bibnamefont
  {Shalizi}}\ and\ \bibinfo {author} {\bibfnamefont {K.~L.}\ \bibnamefont
  {Shalizi}},\ }\href {http://arxiv.org/abs/cs/0406011} {\  (\bibinfo {year}
  {2004})},\ \bibinfo {note} {arXiv: cs/0406011}\BibitemShut {NoStop}%
\bibitem [{\citenamefont {Crutchfield}(2012)}]{crutchfield2012}%
  \BibitemOpen
  \bibfield  {author} {\bibinfo {author} {\bibfnamefont {J.~P.}\ \bibnamefont
  {Crutchfield}},\ }\href {\doibase 10.1038/nphys2190} {\bibfield  {journal}
  {\bibinfo  {journal} {Nature Physics}\ }\textbf {\bibinfo {volume} {8}},\
  \bibinfo {pages} {17} (\bibinfo {year} {2012})},\ \bibinfo {note} {number: 1
  Publisher: Nature Publishing Group}\BibitemShut {NoStop}%
\bibitem [{\citenamefont {Gu}\ \emph {et~al.}(2012)\citenamefont {Gu},
  \citenamefont {Wiesner}, \citenamefont {Rieper},\ and\ \citenamefont
  {Vedral}}]{gu2012}%
  \BibitemOpen
  \bibfield  {author} {\bibinfo {author} {\bibfnamefont {M.}~\bibnamefont
  {Gu}}, \bibinfo {author} {\bibfnamefont {K.}~\bibnamefont {Wiesner}},
  \bibinfo {author} {\bibfnamefont {E.}~\bibnamefont {Rieper}}, \ and\ \bibinfo
  {author} {\bibfnamefont {V.}~\bibnamefont {Vedral}},\ }\href {\doibase
  10.1038/ncomms1761} {\bibfield  {journal} {\bibinfo  {journal} {Nature
  Communications}\ }\textbf {\bibinfo {volume} {3}},\ \bibinfo {pages} {762}
  (\bibinfo {year} {2012})},\ \bibinfo {note} {arXiv: 1102.1994 Publisher:
  Nature Publishing Group}\BibitemShut {NoStop}%
\bibitem [{\citenamefont {Mahoney}\ \emph {et~al.}(2016)\citenamefont
  {Mahoney}, \citenamefont {Aghamohammadi},\ and\ \citenamefont
  {Crutchfield}}]{mahoney2016}%
  \BibitemOpen
  \bibfield  {author} {\bibinfo {author} {\bibfnamefont {J.~R.}\ \bibnamefont
  {Mahoney}}, \bibinfo {author} {\bibfnamefont {C.}~\bibnamefont
  {Aghamohammadi}}, \ and\ \bibinfo {author} {\bibfnamefont {J.~P.}\
  \bibnamefont {Crutchfield}},\ }\href {\doibase 10.1038/srep20495} {\bibfield
  {journal} {\bibinfo  {journal} {Scientific Reports}\ }\textbf {\bibinfo
  {volume} {6}},\ \bibinfo {pages} {20495} (\bibinfo {year} {2016})},\ \bibinfo
  {note} {number: 1 Publisher: Nature Publishing Group}\BibitemShut {NoStop}%
\bibitem [{zot()}]{zotero-1447}%
  \BibitemOpen
  \href@noop {} {\emph {\bibinfo {title} {See supplemental
  material}}}\BibitemShut {NoStop}%
\bibitem [{\citenamefont {Wyner}(1978)}]{wyner1978}%
  \BibitemOpen
  \bibfield  {author} {\bibinfo {author} {\bibfnamefont {A.~D.}\ \bibnamefont
  {Wyner}},\ }\href {\doibase 10.1016/S0019-9958(78)90026-8} {\bibfield
  {journal} {\bibinfo  {journal} {Information and Control}\ }\textbf {\bibinfo
  {volume} {38}},\ \bibinfo {pages} {51} (\bibinfo {year} {1978})},\ \bibinfo
  {note} {iSBN: 0471241954}\BibitemShut {NoStop}%
\bibitem [{\citenamefont {Renner}\ and\ \citenamefont
  {Maurer}(2002)}]{renner2002}%
  \BibitemOpen
  \bibfield  {author} {\bibinfo {author} {\bibfnamefont {R.}~\bibnamefont
  {Renner}}\ and\ \bibinfo {author} {\bibfnamefont {U.}~\bibnamefont
  {Maurer}},\ }\href {\doibase 10.1109/ISIT.2002.1023636} {\bibfield  {journal}
  {\bibinfo  {journal} {IEEE International Symposium on Information Theory}\ ,\
  \bibinfo {pages} {364}} (\bibinfo {year} {2002})},\ \bibinfo {note} {iSBN:
  0-7803-7501-7}\BibitemShut {NoStop}%
\bibitem [{\citenamefont {Cover}\ and\ \citenamefont {J}(1991)}]{cover1991}%
  \BibitemOpen
  \bibfield  {author} {\bibinfo {author} {\bibfnamefont {T.~M.}\ \bibnamefont
  {Cover}}\ and\ \bibinfo {author} {\bibfnamefont {T.~A.}\ \bibnamefont {J}},\
  }\href@noop {} {\emph {\bibinfo {title} {Elements of {Information}
  {Theory}}}}\ (\bibinfo  {publisher} {Wiley},\ \bibinfo {address} {New York},\
  \bibinfo {year} {1991})\BibitemShut {NoStop}%
\bibitem [{\citenamefont {Yang}\ \emph {et~al.}(2020)\citenamefont {Yang},
  \citenamefont {Binder}, \citenamefont {Gu},\ and\ \citenamefont
  {Elliott}}]{yang2020}%
  \BibitemOpen
  \bibfield  {author} {\bibinfo {author} {\bibfnamefont {C.}~\bibnamefont
  {Yang}}, \bibinfo {author} {\bibfnamefont {F.~C.}\ \bibnamefont {Binder}},
  \bibinfo {author} {\bibfnamefont {M.}~\bibnamefont {Gu}}, \ and\ \bibinfo
  {author} {\bibfnamefont {T.~J.}\ \bibnamefont {Elliott}},\ }\href {\doibase
  10.1103/PhysRevE.101.062137} {\bibfield  {journal} {\bibinfo  {journal}
  {Physical Review E}\ }\textbf {\bibinfo {volume} {101}},\ \bibinfo {pages}
  {062137} (\bibinfo {year} {2020})}\BibitemShut {NoStop}%
\bibitem [{\citenamefont {Jurgens}\ and\ \citenamefont
  {Crutchfield}(2021)}]{jurgens2021}%
  \BibitemOpen
  \bibfield  {author} {\bibinfo {author} {\bibfnamefont {A.~M.}\ \bibnamefont
  {Jurgens}}\ and\ \bibinfo {author} {\bibfnamefont {J.~P.}\ \bibnamefont
  {Crutchfield}},\ }\href {\doibase 10.1007/s10955-021-02769-3} {\bibfield
  {journal} {\bibinfo  {journal} {Journal of Statistical Physics}\ }\textbf
  {\bibinfo {volume} {183}},\ \bibinfo {pages} {32} (\bibinfo {year}
  {2021})}\BibitemShut {NoStop}%
\bibitem [{\citenamefont {Travers}\ and\ \citenamefont
  {Crutchfield}(2012)}]{travers2012}%
  \BibitemOpen
  \bibfield  {author} {\bibinfo {author} {\bibfnamefont {N.~F.}\ \bibnamefont
  {Travers}}\ and\ \bibinfo {author} {\bibfnamefont {J.~P.}\ \bibnamefont
  {Crutchfield}},\ }\href {http://arxiv.org/abs/1111.4500} {\enquote {\bibinfo
  {title} {Equivalence of {History} and {Generator} {Epsilon}-{Machines}},}\ }
  (\bibinfo {year} {2012}),\ \bibinfo {note} {arXiv:1111.4500 [cond-mat,
  physics:nlin, stat]}\BibitemShut {NoStop}%
\bibitem [{\citenamefont {van Kampen}(2007)}]{vankampen2007}%
  \BibitemOpen
  \bibfield  {author} {\bibinfo {author} {\bibfnamefont {N.~G.}\ \bibnamefont
  {van Kampen}},\ }\href@noop {} {\emph {\bibinfo {title} {Stochastic
  {Processes} in {Physics} and {Chemistry}}}}\ (\bibinfo  {publisher}
  {North-Holland Personal Library},\ \bibinfo {year} {2007})\BibitemShut
  {NoStop}%
\bibitem [{\citenamefont {Breuer}\ and\ \citenamefont
  {Petruccione}(2007)}]{breuer2007}%
  \BibitemOpen
  \bibfield  {author} {\bibinfo {author} {\bibfnamefont {H.~P.}\ \bibnamefont
  {Breuer}}\ and\ \bibinfo {author} {\bibfnamefont {F.}~\bibnamefont
  {Petruccione}},\ }\href@noop {} {\emph {\bibinfo {title} {The {Theory} of
  {Open} {Quantum} {Systems}}}}\ (\bibinfo  {publisher} {Oxford University
  Press, USA},\ \bibinfo {year} {2007})\BibitemShut {NoStop}%
\bibitem [{\citenamefont {Davies}(1974)}]{davies1974}%
  \BibitemOpen
  \bibfield  {author} {\bibinfo {author} {\bibfnamefont {E.~B.}\ \bibnamefont
  {Davies}},\ }\href {\doibase 10.1007/BF01608389} {\bibfield  {journal}
  {\bibinfo  {journal} {Communications in Mathematical Physics}\ }\textbf
  {\bibinfo {volume} {39}},\ \bibinfo {pages} {91} (\bibinfo {year}
  {1974})}\BibitemShut {NoStop}%
\bibitem [{\citenamefont {González}\ \emph {et~al.}(2017)\citenamefont
  {González}, \citenamefont {Correa}, \citenamefont {Nocerino}, \citenamefont
  {Palao}, \citenamefont {Alonso},\ and\ \citenamefont
  {Adesso}}]{gonzalez2017}%
  \BibitemOpen
  \bibfield  {author} {\bibinfo {author} {\bibfnamefont {J.~O.}\ \bibnamefont
  {González}}, \bibinfo {author} {\bibfnamefont {L.~A.}\ \bibnamefont
  {Correa}}, \bibinfo {author} {\bibfnamefont {G.}~\bibnamefont {Nocerino}},
  \bibinfo {author} {\bibfnamefont {J.~P.}\ \bibnamefont {Palao}}, \bibinfo
  {author} {\bibfnamefont {D.}~\bibnamefont {Alonso}}, \ and\ \bibinfo {author}
  {\bibfnamefont {G.}~\bibnamefont {Adesso}},\ }\href {\doibase
  10.1142/S1230161217400108} {\bibfield  {journal} {\bibinfo  {journal} {Open
  Systems \& Information Dynamics}\ }\textbf {\bibinfo {volume} {24}},\
  \bibinfo {pages} {1740010} (\bibinfo {year} {2017})},\ \bibinfo {note}
  {arXiv: 1707.09228}\BibitemShut {NoStop}%
\bibitem [{\citenamefont {Wichterich}\ \emph {et~al.}(2007)\citenamefont
  {Wichterich}, \citenamefont {Henrich}, \citenamefont {Breuer}, \citenamefont
  {Gemmer},\ and\ \citenamefont {Michel}}]{wichterich2007}%
  \BibitemOpen
  \bibfield  {author} {\bibinfo {author} {\bibfnamefont {H.}~\bibnamefont
  {Wichterich}}, \bibinfo {author} {\bibfnamefont {M.~J.}\ \bibnamefont
  {Henrich}}, \bibinfo {author} {\bibfnamefont {H.~P.}\ \bibnamefont {Breuer}},
  \bibinfo {author} {\bibfnamefont {J.}~\bibnamefont {Gemmer}}, \ and\ \bibinfo
  {author} {\bibfnamefont {M.}~\bibnamefont {Michel}},\ }\href {\doibase
  10.1103/PhysRevE.76.031115} {\bibfield  {journal} {\bibinfo  {journal}
  {Physical Review E}\ }\textbf {\bibinfo {volume} {76}},\ \bibinfo {pages}
  {031115} (\bibinfo {year} {2007})},\ \bibinfo {note} {arXiv:
  quant-ph/0703048}\BibitemShut {NoStop}%
\bibitem [{\citenamefont {Feller}(1949)}]{feller1949}%
  \BibitemOpen
  \bibfield  {author} {\bibinfo {author} {\bibfnamefont {W.}~\bibnamefont
  {Feller}},\ }\href@noop {} {\bibfield  {journal} {\bibinfo  {journal}
  {Transactions of the American Mathematical Society}\ }\textbf {\bibinfo
  {volume} {67.1}},\ \bibinfo {pages} {98} (\bibinfo {year}
  {1949})}\BibitemShut {NoStop}%
\bibitem [{Note1()}]{Note1}%
  \BibitemOpen
  \bibinfo {note} {We could also have used as a distance function the
  trace-distance between the two density matrices. However, we believe that
  clustering based on probabilities is more meaningful as it will bundle
  together states which differ not in their shapes, but in how they affect the
  statistics.}\BibitemShut {Stop}%
\bibitem [{\citenamefont {Fink}\ \emph {et~al.}(2018)\citenamefont {Fink},
  \citenamefont {Schade}, \citenamefont {Hofling}, \citenamefont {Schneider},\
  and\ \citenamefont {Imamoglu}}]{fink2018}%
  \BibitemOpen
  \bibfield  {author} {\bibinfo {author} {\bibfnamefont {T.}~\bibnamefont
  {Fink}}, \bibinfo {author} {\bibfnamefont {A.}~\bibnamefont {Schade}},
  \bibinfo {author} {\bibfnamefont {S.}~\bibnamefont {Hofling}}, \bibinfo
  {author} {\bibfnamefont {C.}~\bibnamefont {Schneider}}, \ and\ \bibinfo
  {author} {\bibfnamefont {A.}~\bibnamefont {Imamoglu}},\ }\href@noop {}
  {\bibfield  {journal} {\bibinfo  {journal} {Nature Physics}\ }\textbf
  {\bibinfo {volume} {14}},\ \bibinfo {pages} {365} (\bibinfo {year} {2018})},\
  \bibinfo {note} {arXiv: 1707.01837}\BibitemShut {NoStop}%
\bibitem [{\citenamefont {Fink}\ \emph {et~al.}(2017)\citenamefont {Fink},
  \citenamefont {Dombi}, \citenamefont {Vukics}, \citenamefont {Wallraff},\
  and\ \citenamefont {Domokos}}]{fink2017}%
  \BibitemOpen
  \bibfield  {author} {\bibinfo {author} {\bibfnamefont {J.~M.}\ \bibnamefont
  {Fink}}, \bibinfo {author} {\bibfnamefont {A.}~\bibnamefont {Dombi}},
  \bibinfo {author} {\bibfnamefont {A.}~\bibnamefont {Vukics}}, \bibinfo
  {author} {\bibfnamefont {A.}~\bibnamefont {Wallraff}}, \ and\ \bibinfo
  {author} {\bibfnamefont {P.}~\bibnamefont {Domokos}},\ }\href {\doibase
  10.1103/PhysRevX.7.011012} {\bibfield  {journal} {\bibinfo  {journal}
  {Physical Review X}\ }\textbf {\bibinfo {volume} {7}},\ \bibinfo {pages}
  {011012} (\bibinfo {year} {2017})},\ \bibinfo {note} {arXiv:
  1607.04892}\BibitemShut {NoStop}%
\end{thebibliography}%

\pagebreak
\widetext

\newpage 
\begin{center}
\vskip0.5cm
{\Large Supplemental Material}
\end{center}
\vskip0.4cm

\setcounter{section}{0}
\setcounter{equation}{0}
\setcounter{figure}{0}
\setcounter{table}{0}
\setcounter{page}{1}
\renewcommand{\theequation}{S\arabic{equation}}
\renewcommand{\thefigure}{S\arabic{figure}}


This supplemental material is divided in two parts. 
In Sec. S1 we prove the results in the main text concerning the spectral properties of the super-operator $\mathcal{M}$. We also prove that the process is CPTP and relate $\mathcal{M}$ with the Drazin inverse of the system Liouvillian. 
In Sec. S2 we provide additional details on the examples discussed in the main text, including how to determine the patterns shown in Figs.~\ref{fig:patterns} and Fig.~\ref{fig:kappa_graphs}.

\section{S1. Spectral properties of $\mathcal{M}$}

\subsection{Proof of stationarity}

Let us consider the general distribution $\mathcal{P}(\tau_1,k_1,\ldots,\tau_N,k_N)$. 
This process will be stationary when 
\begin{equation}
    \mathcal{P}(\tau_1,k_1,\ldots,\tau_N,k_N) = \mathcal{P}(\tau_{i+1},k_{i+1},\ldots,\tau_{i+N},k_{i+N}),\qquad \forall i,N.
\end{equation}
Starting from Eq.~\eqref{P_general} we have 
\begin{equation}
    \mathcal{P}(\tau_1,k_1,\ldots,\tau_{i+N},k_{i+N}) = \tr\big\{ 
    \mathcal{J}_{k_{i+N}} e^{\mathcal{L}_0 \tau_{i+N}} \ldots \mathcal{J}_{k_1} e^{\mathcal{L}_0 \tau_1} \rho \big\}.
\end{equation}
Marginalizing over $1,\ldots,N$ using Eq.~\eqref{replacement_M} gives 
\begin{equation}
    \mathcal{P}(\tau_{i+1},k_{i+1},\ldots,\tau_{i+N},k_{i+N}) = \tr\big\{
        \mathcal{J}_{k_{i+N}}e^{\mathcal{L}_0 \tau_{i+N}}\ldots \mathcal{J}_{k_{i+1}}e^{\mathcal{L}_0 \tau_{i+1}} \mathcal{M}^i \rho
    \big\}.
\end{equation}
This is to be compared with $\mathcal{P}(\tau_1,k_1,\ldots,\tau_N,k_N)$ in Eq.~\eqref{P_general}. 
We see that both will coincide if and only if 
\begin{equation}
    \mathcal{M}^i \rho = \rho, \qquad \forall i, \rho.
\end{equation}
The process will thus be stationary when $\mathcal{M}^i \rho = \rho$.

The existence of a stationary distribution therefore relies on the spectral properties of $\mathcal{M}$. 
Namely, there must exist a state $\pi$ which is a right eigenvector of $\mathcal{M}$ with eigenvalue 1. 
This, it turns out, is given by the  jump steady-state (JSS) in Eq.~\eqref{pi}. 
To verify that this is true, we recall that $\mathcal{L} \rhoss = 0$. 
Then 
\begin{equation}
    \mathcal{M} \pi = -\frac{1}{K} \mathcal{J}\mathcal{L}_0^{-1} \mathcal{J} \rhoss = -\frac{1}{K} \mathcal{J}\mathcal{L}_0^{-1} (\mathcal{L}- \mathcal{L}_0) \rhoss 
    = \frac{1}{K} \mathcal{J} \rhoss = \pi.
\end{equation}
As also mentioned in the main text, because $\mathcal{J} = \mathcal{L}- \mathcal{L}_0$ and since $\tr(\mathcal{L}\bullet) = 0$, it follows that  
\begin{equation}
    \tr\big\{ \mathcal{M}(\bullet)\big\} = -\tr\big\{ (\mathcal{L}- \mathcal{L}_0) \mathcal{L}_0^{-1} \bullet\big\} 
    = \tr\big\{ \bullet\big\}.
\end{equation}
Hence,  $\tr(\bullet)$ is a left eigenvector of $\mathcal{M}$ with eigenvalue $1$.

\subsection{Proof that the process is CPTP}
\label{app:CPTP}

To provide a consistent picture, we start by proving that the map 
\begin{equation}\label{Sup_map_Mk}
    \rho \to \frac{\mathcal{M}_k \rho}{\tr(\mathcal{M}_k\rho)},
\end{equation}
is completely positive and trace preserving (CPTP).
Notice that this is not a linear quantum operation; instead, it is the (non-linear) post-selected outcome of a Kraus channel. Hence, in this sense, the CPTP property is evident. 
Notwithstanding, we provide here with a self-contained proof. 

It suffices to prove that $\mathcal{M}_k \rho$ is CP, since the normalization is embeded in~\eqref{Sup_map_Mk}.
A map of the form $A \rho A^\dagger$ is CP because if $\rho = \sum_k p_k |k\rangle\langle k|$ then 
\begin{equation}\label{app_CPTP_phi_eq}
    \langle \phi | A \rho A^\dagger | \phi \rangle = \sum_k p_k |\langle \phi | A|k\rangle|^2 \geqslant 0, \qquad \forall |\phi\rangle, 
\end{equation}
which is true iff $A\rho A^\dagger \geqslant 0$.

We now use this to solve our actual problem. 
We can write 
\begin{equation}
    \mathcal{M}_k \rho = - \mathcal{J}_k \mathcal{L}_0^{-1} \rho = \int\limits_0^\infty dt ~L_k (e^{\mathcal{L}_0 t} \rho) L_k^\dagger.
\end{equation}
First, let us assume that all channels are monitored. In this case the no-jump operator can be written as 
\begin{equation}
    \mathcal{L}_0 \rho = -i (H_e \rho - \rho H_e^\dagger), 
    \qquad H_e = H - \frac{i}{2} \sum_k L_k^\dagger L_k.
\end{equation}
It then follows that 
\begin{equation}
    \mathcal{M}_k \rho = \int\limits_0^\infty dt~ L_k e^{-i H_e t} \rho e^{i H_e^\dagger t} L_k^\dagger. 
\end{equation}
Using the same argument as in Eq.~\eqref{app_CPTP_phi_eq} we see that 
\begin{equation}
    \langle \phi | \mathcal{M}_k(\rho) | \phi\rangle = \sum_k p_k \int\limits_0^\infty dt~ |\langle \phi | L_k e^{-i H_e t} |k\rangle|^2 \geqslant 0.
\end{equation}
This shows that in the case where all channels are monitored, $\mathcal{M}_k \rho$ is CP. 

Next we extend to the case where only some of the channels are monitored. 
The no-jump operator is defined as $\mathcal{L}_0 = \mathcal{L} - \sum_{k\in \mathbb{M}} \mathcal{J}_k$. 
We also define another no-jump operator for the case where all channels are monitored, $\mathcal{L}_0' = \mathcal{L}- \sum_k \mathcal{J}_k$ (i.e., with the sum being over all channels $k$ and not only $k\in \mathbb{M}$).
It then follows that 
\begin{equation}
    \mathcal{L}_0 = \mathcal{L}_0' + \sum_{k \notin \mathbb{M}} \mathcal{J}_k. 
\end{equation}
We now perform a Dyson series expansion of $e^{\mathcal{L}_0 t}$: 
\begin{equation}
    e^{\mathcal{L}_0 t} \rho = e^{\mathcal{L}_0' t} \rho + \sum_{k \notin \mathbb{M}} \int\limits_0^t dt' ~e {\mathcal{L}_0' (t-t')} \mathcal{J}_k e^{\mathcal{L}_0' t'} \rho + \ldots. 
\end{equation}
We already showed that because $\mathcal{L}_0' \rho = -i (H_e \rho - \rho H_e^\dagger)$, complete positivity is guaranteed for $e^{\mathcal{L}_0' t} \rho$. 
The Dyson series shows, therefore, that the same will be true for each term in the expansion. And since all terms are added with a plus sign, positivity will be ensured by all terms in the sum. 
This therefore concludes the proof that $\mathcal{M}_k \rho$ is CP. 

Since $\mathcal{M}_k \rho$ is a positive operator it follows that $\tr(\mathcal{M}_k\rho) \geqslant 0$. 
And since this can be extended to multiple applications $\mathcal{M}_{k_N}\ldots \mathcal{M}_{k_1}\rho$, it follows that the multi-point probability in Eq.~\eqref{P_channels} is non-negative. 
As we have already shown in the main text that it is properly normalized, then it follows that for any initial state, any Hamiltonian and any set of jump operators, the process in question always generates a valid and proper probability distribution. 

\subsection{Spectrum of $\mathcal{M}$}

We now switch to vectorized notation~\cite{landi2023}. 
We have already shown that 
\begin{equation}
    \mathcal{M} |\pi\rrangle = |\pi\rrangle, 
    \qquad 
    \idV \mathcal{M} = \idV, 
\end{equation}
where $\idV$ is the vectorized identity,  corresponding to the trace. That is, 
$\tr(A) = \idV | A \rrangle$.

In what follows we assume for simplicity that $\mathcal{M}$ is diagonalizable. The same argument holds when it is not, provided we use Jordan forms. 
The remaining eigenvectors/eigenvalues of $\mathcal{M}$ will be denoted by 
\begin{equation}
    \mathcal{M} |u_j\rrangle = \mu_j |u_j\rrangle, 
    \qquad 
    \llangle v_j | \mathcal{M} = \mu_j \llangle v_j|.
\end{equation}
Thus, $\mathcal{M}$ can be decomposed as 
\begin{equation}\label{Sup_M_eigen_decomposition}
    \mathcal{M} = |\pi\rrangle \idV + \sum_j \mu_j |u_j\rrangle\llangle v_j|. 
\end{equation}
The eigenvectors also satisfy the orthonormality condition 
\begin{equation}
    \idV |\pi\rrangle = 1, \qquad 
    \idV |u_j \rrangle = 0, 
    \qquad 
    \llangle v_j | \pi \rrangle = 0, 
    \qquad 
    \llangle v_j |u_k\rrangle = \delta_{j,k}.
\end{equation}

Using these results, we can now show that the eigenvalues of $\mathcal{M}$ cannot be larger than unity. We do that by first deriving Eq.~\eqref{Pk1kN_spectral} of the main text.
We start with Eq.~\eqref{Pk1kN} which describes the joint distribution of two distant emissions. 
In vectorized notation it reads 
\begin{equation}
    \mathcal{P}(k_1,k_N) = \idV \mathcal{M}_{k_N} \mathcal{M}^{N-2} \mathcal{M}_{k_1} |\pi\rrangle. 
\end{equation}
Substituting for Eq.~\eqref{Sup_M_eigen_decomposition} yields 
\begin{equation}
    \mathcal{P}(k_1,k_N) = \idV \mathcal{M}_{k_N} |\pi\rrangle \idV \mathcal{M}_{k_1} |\pi\rrangle + \sum_j \mu_j^{N-2} \idV \mathcal{M}_{k_N} |u_j \rrangle\llangle v_j |\mathcal{M}_{k_1} |\pi\rrangle. 
\end{equation}
The first term is nothing but the single-outcome distribution in Eq.~\eqref{Pk1}. Hence, we conclude that 
\begin{equation}\label{Sup_P_k1kN_eigen}
    \mathcal{P}(k_1,k_N) = \mathcal{P}(k_1) \mathcal{P}(k_N) + \sum_j \mu_j^{N-2} \idV \mathcal{M}_{k_N} |u_j \rrangle\llangle v_j |\mathcal{M}_{k_1} |\pi\rrangle.
\end{equation}
which is Eq.~\eqref{Pk1kN_spectral}, in vectorized notation and $\mathbb{P}_j := |u_j \rrangle\llangle v_j|$. 
From this expression we can conclude that 
\begin{equation}\label{Sup_eigen_M_condition}
    |\mu_j | \leqslant 1. 
\end{equation}
This is because $\mathcal{P}(k_1,k_N) \leqslant 1$, as it is a valid probability distribution. 
And if we had $|\mu_j|>1$ we could always choose a $N$ sufficiently large to make the right-hand side of Eq.~\eqref{Sup_P_k1kN_eigen} larger than unity. 
Since we already showed in the previous section that this is not possible, Eq.~\eqref{Sup_eigen_M_condition} follows. 

\subsection{Relation between $\mathcal{M}$ and the Drazin inverse of the Liouvillian}

We can also relate $\mathcal{M}$ to the Drazin inverse of the original Liouvillian $\mathcal{L}$. 
We assume again that $\mathcal{L}$ is diagonalizable. 
The zero eigenvalue corresponds to the steady-state, with $\mathcal{L} \rhossV = 0$ and $\idV \mathcal{L}  = 0$, where $\rhossV$ is the steady-state in vectorized form and $\idV$ is the trace operation; i.e., $\idV A\rrangle = \tr(A)$. 
The steady-state is assumed to be unique. All other eigenvalues, which we label as $\lambda_j$, therefore have strictly negative real parts. 
We write the corresponding right and left eigenvectors as 
\begin{equation}
    \mathcal{L}|x_j\rrangle = \lambda_j |x_j\rrangle, 
    \qquad 
    \llangle y_j | \mathcal{L} = \lambda_j \llangle y_j |.
\end{equation}
From this it then follows that $\mathcal{L} = \sum_j \lambda_j |x_j\rrangle\llangle y_j|$. 
The Drazin pseudo-inverse is defined as 
\begin{equation}
    \mathcal{L}^+ = \sum_j \frac{1}{\lambda_j}|x_j\rrangle\llangle y_j|.
\end{equation}

We start by relating $\mathcal{L}_0^{-1}$ with $\mathcal{L}^+$, where $\mathcal{L}_0 = \mathcal{L}- \mathcal{J}$. 
We do this using the Sherman-Morrison-Woodbury (SMW) formula
\begin{equation}\label{SMW}
    (A+B)^{-1} = A^{-1} - A^{-1} (A^{-1} + B^{-1})^{-1} A^{-1},
\end{equation}
which can also be rearranged as 
\begin{equation}\label{SMW2}
    (A+B)^{-1} = A^{-1} (B A^{-1} + 1)^{-1}.
\end{equation}
Neither $\mathcal{L}$ nor $\mathcal{J}$ are invertible, however. So we introduce an infinitesimal parameter $\epsilon$ and write instead $\mathcal{L}_0 = \mathcal{L}_\epsilon - \mathcal{J}_\epsilon$, where $\mathcal{L}_\epsilon = \mathcal{L} + \epsilon$ and similarly for $\mathcal{J}_\epsilon$. 
In the end, we take the limit $\epsilon\to 0$. 
Eq.~\eqref{SMW2} now yields 
\begin{equation}\label{Loinv_step1}
    \mathcal{L}_0^{-1} = - \mathcal{L}_\epsilon^{-1} \big( \mathcal{J}_\epsilon \mathcal{L}_\epsilon^{-1}-1)^{-1}.
\end{equation}
Moreover,
\begin{equation}
    \mathcal{L}_\epsilon^{-1} = \frac{1}{\epsilon} \rhossV \idV + \sum_j \frac{1}{\lambda_j + \epsilon} |x_j\rrangle\llangle y_j|.
\end{equation}
Since all $\lambda_j$ have strictly negative real parts, and since we are only interested in the limit $\epsilon\to 0$, we can approximate the last term by $\frac{1}{\lambda_j + \epsilon} \simeq \frac{1}{\lambda_j}$.
We therefore arrive at 
\begin{equation}\label{L_epsilon_inverse}
    \mathcal{L}_\epsilon^{-1} = \frac{1}{\epsilon} \rhossV \idV + \mathcal{L}^+.
\end{equation}

Referring to Eq.~\eqref{Loinv_step1}, what we need is to find the inverse of
\begin{equation}
    \mathcal{J}_\epsilon \mathcal{L}_\epsilon^{-1} - 1 =  \mathcal{J}_\epsilon\mathcal{L}^+ - 1 + \frac{1}{\epsilon}\mathcal{J}_\epsilon \rhossV \idV.     
\end{equation}
To leading order in $\epsilon$, we are allowed to replace $\mathcal{J}_\epsilon$ with $\mathcal{J}$, so that the matrix in question becomes 
\begin{equation}
    \mathcal{J}_\epsilon \mathcal{L}_\epsilon^{-1} - 1= \mathcal{J}\mathcal{L}^+ -1 + \frac{1}{\epsilon}\mathcal{J} \rhossV \idV.     
\end{equation}
This can be inverted using the identity 
\begin{equation}\label{rank_1_SMW}
    (A + |u\rrangle\llangle v |)^{-1} = A^{-1} - \frac{A^{-1} |u\rrangle\llangle v | A^{-1}}{1 + \llangle v | A^{-1} | u\rrangle.},
\end{equation}
which yields 
\begin{equation}
    (\mathcal{J}_\epsilon \mathcal{L}_\epsilon^{-1}-1)^{-1} = B - 
    \frac{
        B \mathcal{J} \rhossV \idV B
    }{
        \epsilon + g
    },
\end{equation}
where $B = (\mathcal{J}\mathcal{L}^+-1)^{-1}$ and $g=\idV B \mathcal{J} \rhossV$ are just shorthand notations.
We can now plug this back into Eq.~\eqref{Loinv_step1}, together with Eq.~\eqref{L_epsilon_inverse}. 
As a result, we find that 
\begin{equation}
    \mathcal{L}_0^{-1} = - \frac{\rhossV \idV B}{\epsilon+g} - \mathcal{L}^+ B + \frac{\mathcal{L}^+ B \mathcal{J} \rhossV \idV B}{\epsilon+ g}.
\end{equation}
We can now trivially take the limit $\epsilon\to 0$, which finally leads, after a bit of rearranging, to
\begin{equation}\label{L0_drazin_relation}
    \mathcal{L}_0^{-1} =  - \mathcal{L}^+ B + \frac{1}{g}\Big(\mathcal{L}^+ B \mathcal{J} -1\Big)\rhossV \idV B.
\end{equation}
This is the desired relation between $\mathcal{L}_0^{-1}$ and $\mathcal{L}^+$. 
We also note in passing that one may write 
\begin{equation}
    B^{-1} = \mathcal{J} \mathcal{L}^+ - 1 = - \rhossV \idV - \mathcal{L}_0 \mathcal{L}^+.
\end{equation}

Next we consider the matrix $\mathcal{M} = - \mathcal{J} \mathcal{J}_0^{-1}$ which, as discussed in the main text, dictates the memory of the Markov process. 
Multiplying Eq.~\eqref{L0_drazin_relation} by $\mathcal{J}$and noting that $\mathcal{J}\mathcal{L}^+ = 1+B^{-1}$ yields 
\begin{equation}\label{M_drazin_relation}
    \mathcal{M} = 1 + B - \frac{B \mathcal{J} \rhossV \idV B}{g}.
\end{equation}

\subsection{Connection with Hidden-Markov models}

In vectorized notation the $N$-clicks distribution~\eqref{P_channels} reads 
\begin{equation}\label{Sup_P_channels_vectorized}
    P(k_1,\ldots,k_N) = \idV \mathcal{M}_{k_N}   \ldots \mathcal{M}_{k_1} |\pi\rrangle.
\end{equation}
We will now discuss the similarity and differences between this and a hidden Markov model (HMM). 
The HMM is described by a series of hidden states $s_i$, running over some alphabet $s_i \in \mathcal{A}$ and a series of emitted symbols $k_i \in \mathcal{M}$. 
The $N$-point transition probability is given by 
\begin{equation}\label{Sup_HMM}
    P(k_1,\ldots,k_N) = \sum_{s_0,\ldots,s_N} P(k_N,s_N|s_{N-1}) \ldots P(k_1,s_1|s_0) \pi_{s_0},
\end{equation}
where $P(k,s|s')$ is the probability that the hidden layer transitions from $s'\to s$ and a symbol $k$ is emitted in the process. 
The hidden layer evolves according to a Markov chain with transition matrix $P(s|s') = \sum_k P(k,s|s')$. 
Moreover, $\pi_s$ is the steady-state of this dynamics, $\sum_s' P(s|s') \pi_{s'} = \pi_s$. 
We can interpret the transition probabilities $P(k,s|s')$ as a set of $|\mathbb{M}|$ matrices $M_k$ with entries 
\begin{equation}
    (M_k)_{s,s'} = P(k,s|s').
\end{equation}
Then Eq.~\eqref{Sup_HMM} is rewritten as
\begin{equation}\label{Sup_HMM2}
    P(k_1,\ldots,k_N) = \langle 1 | M_{k_N} \ldots M_{k_1} |\pi\rangle, 
\end{equation}
where $\langle 1|$ is a vector with all ones and $|\pi\rangle$ is the vector with entries $\pi_\sigma$. 

The similarity with Eq.~\eqref{Sup_P_channels_vectorized} is now evident. 
Notwithstanding, there are fundamental differences, which is what makes the quantum process interesting. 
First, in Eq.~\eqref{Sup_HMM2} the matrices $M_k$ have all non-negative entries since they correspond to probabilities. 
The same is not true for $\mathcal{M}_k$, which can have negative or complex entries. 
Second, the set of states that the HMM runs is a finite alphabet $s \in \mathcal{A}$.
Conversely, the matrices $\mathcal{M}_k$ live in a space spanned by the basis $|i\rangle\langle j|$ of the Hilbert space. 
Thus, while any $s$ is a physical state in an HMM, in the quantum case only diagonal entries $|i\rangle\langle i|$ will be. 
This shows that in the case of a purely incoherent dynamics the transitions in question become a hidden Markov model. But whenever coherences are present, this is no longer the case. 

\section{S2. Determining patterns in the XX and XY chains}

\subsection{XX chain ($\kappa = 0$)}

Here we discuss how to determine whether Eq.~\eqref{rho_dynamics} supports a closed pattern, in the sense of Eq.~\eqref{pattern}, or at least recurring states.
Visualizing density matrices is not easy. 
One could plot specific observables, such as the population.
But there is no single observable which properly captures all features of a density matrix.
To do it, we proceed as follows. 
We use infinite-precision arithmetic to evolve the system under Eq.~\eqref{rho_dynamics}. 
This means that all superoperators are expressed in terms of exact rational numbers. 
Then, as the system evolves, we assign a new integer label $1,2,\ldots$ every time the density matrix is different from all others that appeared before it.
That is, the initial state $\pi$ is labeled as $1$. 
If $\mathcal{M}_{k_1}\pi/\tr(\mathcal{M}_{k_1} \pi) \neq \pi$ we label it as 2. And so on. 
Due to infinite precision, two density matrices are deemed equal if all its entries are exactly the same (there is no approximation in comparing them). 
By proceeding in this way, we can know whether during the dynamics there are any recurring states; that is, whether the density matrix ever repeats. 

\begin{figure}
    \centering
    \includegraphics[width=0.9\textwidth]{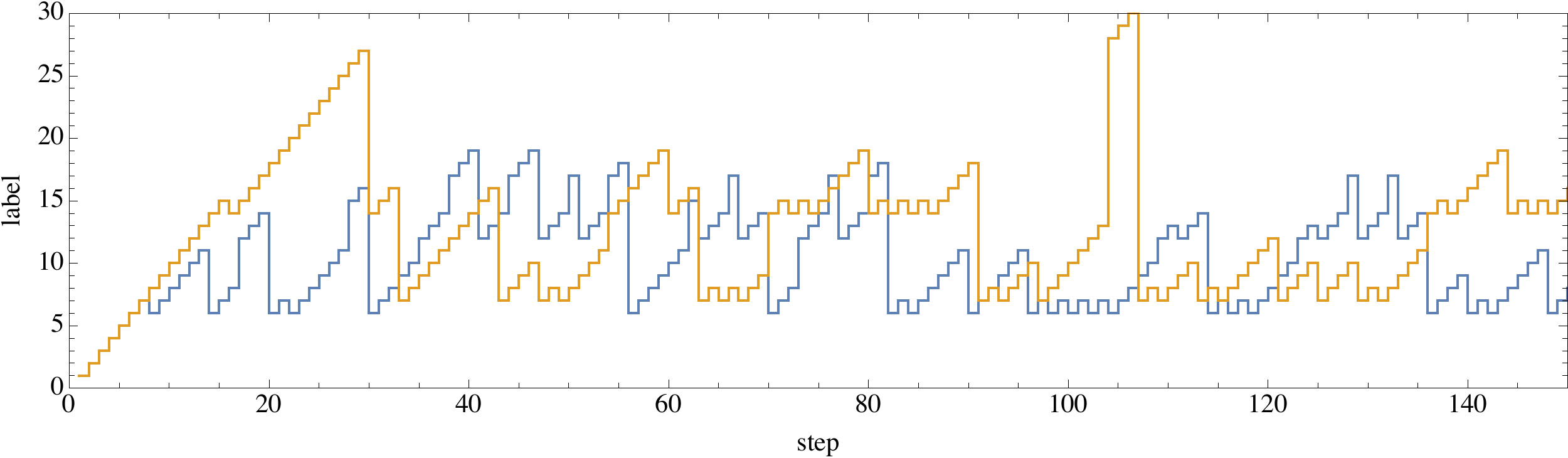}
    \caption{Example of two stochastic trajectories for $L=3$, $\gamma/J = 1$ and $\kappa = 0$ (no pairing terms), showing the label of the density matrix as a function of the step, as the system evolves according to Eq.~\eqref{rho_dynamics}. }
    \label{fig:pattern_label_kappa0}
\end{figure}

Fig.~\ref{fig:pattern_label_kappa0} illustrates the result with two stochastic trajectories for the XX example, with $L=3$, $\gamma/J=1$ and $\kappa = 0$ (that is, no pairing terms).
As can be seen, starting at $\pi$,  there is a series of transient states that the system samples. 
However, after some time, the states start to repeat. 
If a state ever repeats, then we can use that as the seed to determine the pattern by investigating all possible branches.
That is, labeling the repeating state as $\sigma_1$, we can then simply compute 
\begin{equation}
    \frac{\mathcal{M}_{k_n}\ldots \mathcal{M}_{k_1} \sigma_1}{\tr\big\{ \mathcal{M}_{k_n}\ldots \mathcal{M}_{k_1} \sigma_1\big\}},
\end{equation}
and see which classes of states are sampled. 

For $L=3$ and $\kappa = 0$ we obtain the pattern graph in Fig.~\ref{fig:patterns}(c) in the main plot. 
Starting from $|111\rangle$ one can check that the only possible transition is to $|110\rangle$. 
Then, as explained in the main text, 
from $|110\rangle$ it either resets back to $|111\rangle$ via $I$ or it will go to a non-trivial single-particle state $\sigma_{\rm 1p}^{(1)}$, which in the standard computational basis reads 
\begin{equation}
    \sigma_{\rm 1p}^{(1)}= \left(
\begin{array}{cccccccc}
 0 & 0 & 0 & 0 & 0 & 0 & 0 & 0 \\
 0 & 0 & 0 & 0 & 0 & 0 & 0 & 0 \\
 0 & 0 & 0 & 0 & 0 & 0 & 0 & 0 \\
 0 & 0 & 0 & 1-\frac{3}{3 \gamma ^2+5} & 0 & \frac{3 i \gamma }{3 \gamma ^2+5} & 0 & 0 \\
 0 & 0 & 0 & 0 & 0 & 0 & 0 & 0 \\
 0 & 0 & 0 & -\frac{3 i \gamma }{3 \gamma ^2+5} & 0 & \frac{3}{3 \gamma ^2+5} & 0 & 0 \\
 0 & 0 & 0 & 0 & 0 & 0 & 0 & 0 \\
 0 & 0 & 0 & 0 & 0 & 0 & 0 & 0 \\
\end{array}
\right).
\end{equation}
From this it can either go to $|000\rangle$ via $E$ or go to a 2-particle state
\begin{equation}
    \sigma_{\rm 2p}^{(2)}=\left(
\begin{array}{cccccccc}
 0 & 0 & 0 & 0 & 0 & 0 & 0 & 0 \\
 0 & 1-\frac{3 \left(3 \gamma ^2+4\right)}{9 \gamma ^4+30 \gamma ^2+28} & \frac{3 i \gamma  \left(3 \gamma
   ^2+4\right)}{9 \gamma ^4+30 \gamma ^2+28} & 0 & 0 & 0 & 0 & 0 \\
 0 & -\frac{3 i \gamma  \left(3 \gamma ^2+4\right)}{9 \gamma ^4+30 \gamma ^2+28} & \frac{9 \gamma ^2+12}{9
   \gamma ^4+30 \gamma ^2+28} & 0 & 0 & 0 & 0 & 0 \\
 0 & 0 & 0 & 0 & 0 & 0 & 0 & 0 \\
 0 & 0 & 0 & 0 & 0 & 0 & 0 & 0 \\
 0 & 0 & 0 & 0 & 0 & 0 & 0 & 0 \\
 0 & 0 & 0 & 0 & 0 & 0 & 0 & 0 \\
 0 & 0 & 0 & 0 & 0 & 0 & 0 & 0 \\
\end{array}
\right).
\end{equation}
And from this, it can reset to $|111\rangle$ via $I$ or go back to the 1-particle subspace, in yet another new state 
\begin{equation}
    \sigma_{\rm 1p}^{(3)}=\left(
\begin{array}{cccccccc}
 0 & 0 & 0 & 0 & 0 & 0 & 0 & 0 \\
 0 & 0 & 0 & 0 & 0 & 0 & 0 & 0 \\
 0 & 0 & 0 & 0 & 0 & 0 & 0 & 0 \\
 0 & 0 & 0 & \frac{27 \left(\gamma ^4+4 \gamma ^2+6\right) \gamma ^2+80}{27 \left(\gamma ^4+5 \gamma
   ^2+9\right) \gamma ^2+152} & 0 & \frac{9 i \gamma  \left(3 \gamma ^4+9 \gamma ^2+8\right)}{27 \left(\gamma
   ^4+5 \gamma ^2+9\right) \gamma ^2+152} & 0 & 0 \\
 0 & 0 & 0 & 0 & 0 & 0 & 0 & 0 \\
 0 & 0 & 0 & -\frac{9 i \gamma  \left(3 \gamma ^4+9 \gamma ^2+8\right)}{27 \left(\gamma ^4+5 \gamma
   ^2+9\right) \gamma ^2+152} & 0 & \frac{9 \left(3 \gamma ^4+9 \gamma ^2+8\right)}{27 \left(\gamma ^4+5
   \gamma ^2+9\right) \gamma ^2+152} & 0 & 0 \\
 0 & 0 & 0 & 0 & 0 & 0 & 0 & 0 \\
 0 & 0 & 0 & 0 & 0 & 0 & 0 & 0 \\
\end{array}
\right).
\end{equation}
As this example illustrates, as long as the system keeps bouncing back and forth in $IEIE\ldots$ it will continue to sample new states which never repeat. 

\subsection{XY chain ($\kappa \neq 0$)}

\begin{figure}
    \centering
    \includegraphics[width=0.5\textwidth]{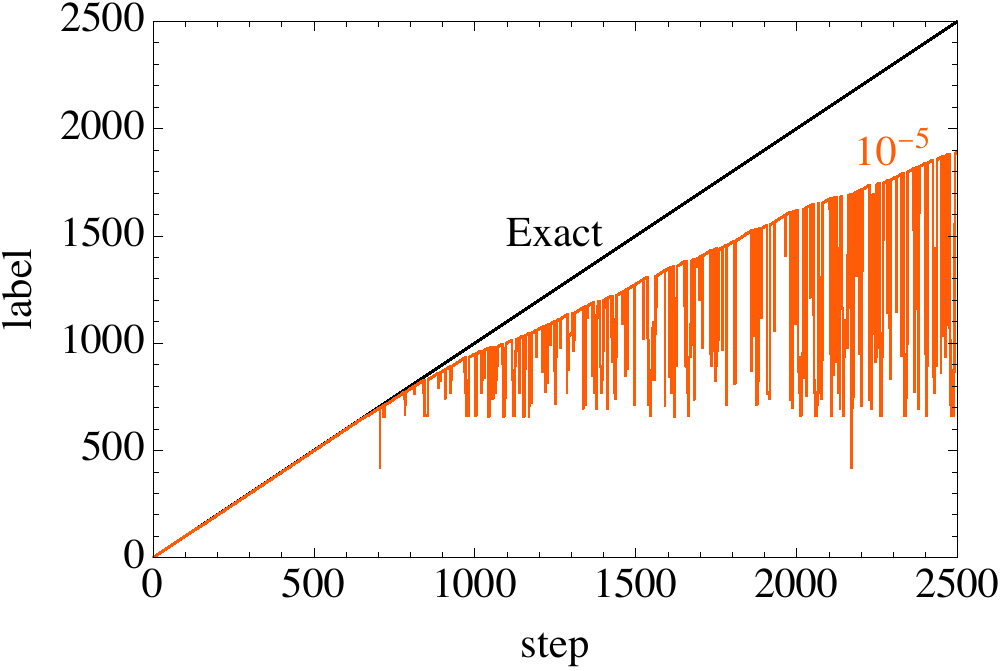}
    \caption{A stochastic trajectory of labels for $L=3$, $\gamma/J = 1$ and $\kappa/J = 1/2$. 
    The black curve is for exact comparison between two density matrices, showing that strictly speaking the states never repeat. 
    The red curve is similar, but compare density matrices up to a certain tolerance $10^{-5}$ in the trace distance between two matrices. 
    This shows that even with a certain tolerance the number of states still continues to grow. 
    }
    \label{fig:pattern_label_kappa_half}
\end{figure}

Next we turn to the case $\kappa \neq 0$. 
We first show that in this case the states actually never repeat. 
In Fig.~\ref{fig:pattern_label_kappa_half} we show a stochastic trajectory for $L=3$, $\gamma/J = 1$ and $\kappa/J = 1/2$.
The black curve is 
similar in spirit to Fig.~\ref{fig:pattern_label_kappa0}.
Now, however, we get a straight line meaning  states never repeat. 
We also show a similar plot, but using an approximate instead of exact comparison. 
That is, we say a density matrix is a new element if it differs from all others up to some tolerance. 
As distance function, we use the trace distance. 
The plots show that even using fairly large tolerances, the number of states still continues to increase. 

\end{document}